\documentclass[12pt]{article}
\usepackage{setspace}
\usepackage[table]{xcolor}
\usepackage[margin=1.15 in]{geometry}
\usepackage{authblk}
\usepackage{amsthm,amsmath,amssymb}
\usepackage{cases}
\usepackage{dsfont}
\usepackage{multirow}
\usepackage{float}

\usepackage{dsfont}

\numberwithin{equation}{section}

\newtheorem{theorem}{Theorem}[section]
\newtheorem{corollary}{Corollary}[section]

\newtheorem{remark}{Remark}[section]

\def\E{\mathbb{E}}
\def\V{\textrm{Var}}

\numberwithin{equation}{section}
\date{}

\begin{document}
\title{Confidence intervals with higher accuracy for short and long memory linear processes}



\author{Masoud M. Nasari         and         Mohamedou Ould-Haye 
\\
\small{School of Mathematics and Statistics of Carleton University}\\ \small{ Ottawa, ON, Canada}
}

\maketitle

\begin{abstract}
\noindent
In this paper  an easy  to implement method of stochastically  weighing short and long memory linear processes is introduced. The method renders asymptotically exact size  confidence intervals for the population mean which are significantly more accurate than  their classical counterparts for each fixed sample size $n$.
It is illustrated both theoretically and numerically that the  randomization framework of this paper  produces randomized (asymptotic) pivotal quantities, for the mean, which admit central limit theorems with smaller magnitudes of error as compared to those of their leading classical counterparts.
An Edgeworth expansion result for randomly weighted linear processes whose   innovations  do not necessarily satisfy the Cramer condition,     is also established.

\end{abstract}

\section{Introduction}\label{On the Skewness of non-iid stationary data}
In a nonparametric framework, the  central limit theorem (CLT)  is a critical  tool in drawing inference about the  mean of a  population. The CLT validates   the use of the percentiles of the normal distribution to approximate those of the unknown sampling distribution of a centered partial sum of a given set of data.
The next  important step after establishing a CLT  is to characterize the  departure of the sampling distribution of the underlying partial sum  functional from normality and to address the  speed at which it  vanishes. Expressed as a function of the sample size $n$, and without restricting the distribution of the data to the symmetrical ones, the error of the CLT   is  generally known to be  of order $O(1/\sqrt{n})$.  Usually, the vanishing rate $1/\sqrt{n}$  is referred to as the first order accuracy, correctness or efficiency of the CLT. The error of the CLT has been extensively studied by several authors, mainly  using the Berry-Esseen inequality and  the Edgeworth expansion. Both these  methods  show that in the i.i.d case, the CLT is usually  first order efficient (cf., e.g.,   Bhattacharya and Rao \cite{Bhattacharya and Rao}, Senatov \cite{Senatov} and Shorack \cite{Shorack}). The same error rate of $1/\sqrt{n}$ has also been shown to hold for some classes of weakly dependent data   (cf., for example,  G\"{o}tze and Hipp \cite{Gotze and Hipp}, Lahiri \cite{Lahiri 1993}).
\\
There are a number of methods in the literature that deal with the problem of how  to increase  the accuracy of the CLT. The tilted approximation  (cf. Barndorff-Nielson and Cox \cite{Barndroff-Nielson and Cox}) is a method of reducing the error admitted by the CLT. In this method, which is of theoretical interest, the densities of  the partial sums of an i.i.d. data set  are  approximated by some tilting  expressions.  The error of this approximation is of order $O(1/n)$.  The drawback of the tilted approximation  is that computing the tilting expressions    requires  the  full  knowledge of the cumulant generating function of the data which is usually unknown.
Also, in the context of i.i.d. data, some other error reduction techniques are built around  adjusting   the cut-off points,   the usual percentiles of standard normal distribution,  by some additive correcting factors (cf., for example, Hall  \cite{Hall Edgeworth correction}). This method is called the Edgeworth correction and it requires estimating   values of  $r$-th moments   of the data, where $r \leq k$, for some $k \geq 3$. The drawback of this method is that it tends to over-correct   the coverage probability of the resulting confidence intervals, when the sample size is relatively small. This issue is the result of the significant deviation between the actual values of the moments of the population  from their estimated values, when the sample is of small size.  Another Edgeworth correction method was introduced in Eriksson \cite{Eriksson}. The approach is essentially a randomization via adding a simulated set of values, from a specific class of distributions, to the original data. To specify the class of distributions from which the additive values are to be generated, it is assumed that the values of the second and third moments   of the original data  are known. Furthermore,  Eriksson \cite{Eriksson} provides   only one example of the distributions which can be used  to generate the additive random values from. The approach is interesting in general, however,  it is unsuitable for use in a nonparametric framework.
The bootstrap (cf., for example, Efron \cite{Efron}) is another approach  to reducing  the error of the CLT that   requires  repeated re-sampling from a given set of data and, hence  it can be intrinsically  viewed as randomized jackknife. Despite its popularity and advantages, the bootstrap approach has  some drawbacks.  This approach relies on repeatedly re-sampling with replacement from a given data set  which makes it computationally expensive. Furthermore, the use of the bootstrap in application poses the important question of how large the number of  bootstrap replications, $B$, should be in order to achieve a higher accuracy for the CLT in contrast to its classical counterpart. While the general belief is that for achieving the goal of higher accuracy,  ideally  $B$ should be $\infty$ (cf. Yatracos \cite{Yatracos}), some suggest that $B=500$ to $B=1000$ suffices (cf. Efron and Tibshirani \cite{Efron and Tibshirani}), others suggest that the higher accuracy is available  uniformly in $B$. This means that the higher accuracy can  still be achieved with replications  as small as  $B=9$ or $B=19$, for instance,   (cf. Hall \cite{Hall}).
It seems that there are no  results in the literature that would relate  the required number of bootstrap replications, $B$,   to the size of the sample at hand. Furthermore, when bootstrap is to be put in use for   dependent data, via  re-sampling with replacement, the i.i.d. nature of the bootstrap samples fails to preserve  the covariance structure of the original data (cf. Cs\"{o}rg\H{o} \textit{et al}. \cite{Csorgo and Nasari and Oul-Haye} for discussions on the effect of bootstrapping on the covariance structure of linear processes). Overcoming this issue requires some adjusted re-sampling schemes, such as over-lapping and non-overlapping block-bootstrapping techniques  (cf., e.g., Lahiri  \cite{Lahiri Resampling Book} and Kim and Nordman   \cite{Kim and Nordman}, G\"{o}tze and K\"{u}nsch  \cite{Gotze and Kunch}), and the sieve   and the augmented sieve (cf., e. g., Poskitt \emph{et al.}    \cite{Pos}). When dealing with long memory data these various adapted for dependant data versions of the bootstrap suffer from a significant underestimation of the variance  which in turn translates in producing confidence intervals with poor coverage probability  even for samples with moderate sizes (cf.  numerical studies in Section \ref{Comparison to bootstrap confidence intervals  for long memory linear processes}).    %
\\
In this paper we introduce a method to stochastically weighing a given set of  data in a multiplicative way  that  results in more accurate CLTs for  linear processes. This approach creates a framework that requires neither  re-sampling nor extra knowledge on the distribution of the data except for the same  commonly assumed moment conditions for the   classical CLT.
\\
The idea of boosting the accuracy of   the CLT via creating  randomized versions of  the  (asymptotic)  pivots of interest, that continue to possess the pivotal property for the parameter of interest,  was first explored in Cs\"{o}rg\H{o} and Nasari \cite{Csorgo and Nasari} for i.i.d. data  and in Cs\"{o}rg\H{o} \textit{et al.} \cite{Csorgo and Nasari and Oul-Haye} for  linear processes. The viewpoint in the latter papers is to create pivots that admit CLTs with smaller magnitude  of error by replacing a data set by a randomly weighted version  of it, where the multiplicative weights are of the form of functionals of  symmetric  multinomial random variables. A broader randomization framework for i.i.d. data was introduced in Nasari \cite{Nasari},  which is flexible in the sense that it allows choosing the randomizing weights from a virtually unlimited class of random weights, including multinomial weights, with a window parameter to regulate the trade-off between the accuracy of the CLT and the volume  of the resulting  randomized confidence regions for the mean of the original data. The present paper constitutes a  generalization of the randomization approach introduced in Nasari \cite{Nasari} for i.i.d. data  to the case when the data form a  short or long memory linear process.
\\
As will be seen in this paper, introducing a controlled  extra source of randomness  in conjunction  with the random mechanism that produces the original data can  enhance the accuracy of the CLT. The refinement in the accuracy results from multiplying the original data by appropriately chosen random factors which do not change the nature of the given data set. More precisely, the multiplicative stochastic weights used in this paper are so that, if the original data set is of short memory then so is their  randomized version. The same continues to hold true  for randomized long memory linear processes. Moreover, in the   case of long memory linear processes, the  randomized pivots introduced in  this  paper tend to yield  significantly better probabilities of coverage as compared to a number of the  existing bootstrap methods including the sieve and the augmented sieve (cf. Section \ref{Comparison to bootstrap confidence intervals  for long memory linear processes}).
\\
We conclude this section by an outline of the topics addressed in the rest of the paper.
In Section \ref{skewness reduction} we introduce  a class of randomized pivotal quantities for the value of the common  mean of the population from which a  linear process structured data set of size $n\geq 1$ is drawn.  In Section \ref{EE Section} we  derive a general Edgeworth expansion result for a class of  randomly weighted linear processes which, in the univariate case, is  an extension of the Edgeworth expansion obtained in  G\"{o}tze and Hipp \cite{Gotze and Hipp} for sums of weakly dependent random vectors. This result, that is also of independent interest, is then used to illustrate the error reduction effect of the  randomization framework in  Section \ref{skewness reduction} on the CLT of the randomized pivots introduced in it.
Section \ref{The CLT} is devoted to establishing the asymptotic  validity  of the randomized pivots introduced in Section \ref{skewness reduction} via deriving a CLT  for them under more relaxed conditions than those required for the Edgeworth expansion result in Section \ref{EE Section}. Section \ref{The CLT} also contains some discussions on the resulting randomized confidence intervals and some numerical demonstrations of their improved performance over that of  their classical  non-randomized counterparts. A numerical comparison of the performances of some methods of bootstrap and the randomization scheme introduced in this paper is presented in Section \ref{Comparison to bootstrap confidence intervals  for long memory linear processes} for long memory data.
Further results  on Studentization of the randomized pivotal quantities are given  in Section  \ref{Complete Studentization}.   Section \ref{proofs} includes the proofs.

\section{Skewness reduction }\label{skewness reduction}
In this section we discuss the shortcoming of the leading measures of skewness which are defined primarily for   i.i.d. data, when dealing with  manipulations (functionals) of dependent data. Then, we consider  a  generalization of the criterion of skewness so that it is suitable for both stationary and i.i.d. data. This generalized definition of skewness is  then used to introduce our approach to increase the  accuracy of the CLT for linear processes.
\\
When the sampling distribution of the normalized partial sums of a set of $n$, $n \geq 1,$ univariate i.i.d. random variables $Y_1, \ldots, Y_n$, admits an  Edgeworth expansion, the Fisher-Pearson's measure of skewness $\E (Y_1-\E(Y_1) )^3 / (\textrm{Var}(Y_1) )^{3/2}$   becomes  the coefficient of an expression of order    $O(1/\sqrt{n})$, as $n \to \infty$. In other words, the measure of skewness is the coefficient of the slowest vanishing term of the error admitted by the CLT. This property motivates viewing the skewness as the most important  characteristic of the distribution of the data in measuring  the  departure from normality.
The closer the value of the skewness is to zero, the more symmetrical the sampling distribution of the underlying partial sum of a given data set will be. The commonly used measures of skewness, such as the aforementioned   Fisher-Pearson,  are defined for the marginal distribution of an i.i.d. set of data  with a finite third moment.  This definition   is   suitable only for the i.i.d. case, as it  does not account for the dependence when  a  stationary and dependent data of size $n$ form  the summands of a partial sum which admits the CLT. The latter observation calls for a broader  definition for  skewness that is to take  the partial sum of a set of random variables into consideration. This means that  skewness should naturally be defined  for the sampling distribution of an underlying  partial sum functional rather than   for the marginal distribution of its summands. The need for such a definition for skewness in the context of this paper stemmed from our need   for a proper measure of skewness  when studying the problem of  boosting  the accuracy of  pivotal quantities  of the form of partial sums  for the mean of linear processes.
Prior to defining an appropriate measure of skewness for stationary linear processes, we first set up the definition of linear processes of consideration.
To formally state our results in this section, and also  for the use throughout this paper, we let  $X_1,\ldots,X_n$ be the first $n\geq 1$  terms of the linear process
$\{X_t:\ t\ge1\}$ defined as

\begin{equation}\label{eq 1}
X_t=\mu+\sum_{k=0}^{\infty} a_{k} \zeta_{t-k}, \ t \geq 1,
\end{equation}
where $\mu$, $\mu \in \mathbb{R}$,  is the mean of the process, $\{a_k; \ k \in \mathbb{Z}  \}$ is a sequence of real numbers such that $\sum_{k=0}^{\infty} a^{2}_{k}<\infty$ and $\{\zeta_k:\ k \in \mathbb{Z} \}$  are i.i.d. white noise innovations  with $\E \zeta_1=0$ and $0<\sigma^2:=\V(\zeta_1)<\infty$.  Moreover,  we assume that      $X_t$ are non-degenerate with  a  finite variance    $\gamma_{0}:= \E X_{1}^2-\mu^2:=\sigma^{2} \sum_{k=0}^{\infty} a^{2}_{k}-\mu^2$ and the autocovariance function

\begin{equation}\label{eq 1'}
\gamma_h:= \textrm{Cov}(X_s,X_{s+h})=\mathbb{E}[(X_{s}-\mu)(X_{s+h}-\mu)], \ h\geq 0, \ s\geq 1.
\end{equation}
In this paper we consider two types of the linear process  (\ref{eq 1}). $(i)$ when  $\sum_{k=0}^{\infty} |a_{k}|= \infty$. In this case   we refer to $X_t$ as a long memory linear  process. In particular, we consider the case when, as $k \rightarrow \infty$, for  some $c>0$ we have $a_k \sim c k^{d-1}$, where $0< d <1/2$. We refer to $d$ as the memory parameter. The other type of linear  processes considered in this paper is $(ii)$ when $\sum_{k=0}^{\infty} |a_{k}|< \infty$. In this case  we refer to $X_t$ as a short memory linear process. To unify our notation we  define the memory parameter   $d$ for short memory linear processes as $d=0$.
\\
Consider the classical pivotal quantity for $\mu$, $T_{n,X}:\mathbb{R}^{n+1}\to\mathbb{R}$,  that is a  normalized partial sum functional, defined as

\begin{equation}\label{T_{n,X}}
 T_{n,X}= \sum_{i=1}^n(X_i-\mu) \big/ \big(\textrm{Var}(\sum_{i=1}^n X_i) \big)^{1/2}   .
\end{equation}
When  $\E|X_1|^3<\infty$, we define the skewness measure  of the pivotal quantity $T_{n,X}$ as follows:
\begin{equation}\label{skw}
\beta_{T_{n,X}}:=\mathbb{E}(T_{n,X})^3 = \E\big(\sum_{i=1}^n(X_i-\mu)\big)^3  \big/
\big(\V (\sum_{i=1}^n X_i )\big)^{3/2}.
\end{equation}
In view of the stationarity of $X_t$, and  without loss of generality,   we assume that $\mu=0$ and expand the  skewness measure $\beta_{T_{n,X}}$   as follows:

\begin{eqnarray}\label{skewness4}
\beta_{T_{n,X}} &=& \E\Big( \sum_{i=1}^n X_i \Big)^3 \big/ \Big( \V (\sum_{i=1}^n X_i ) \Big)^{3/2}\nonumber\\
&=& \E(X_1^3)\big/ \Big( \sqrt{n}( \gamma_0+2\sum_{h=1}^n\big(1-\frac{h}{n}\big)\gamma_h )^{3/2}\Big)\nonumber\\
&+& 3 \Big( \sum_{h=1}^n (1-\frac{h}{n} )\big(\E(X_1^2X_{1+h})+\E(X_1X_{1+h}^2)\big) \Big) \big/ \Big( \sqrt{n}( \gamma_0+2\sum_{h=1}^n (1-\frac{h}{n} )\gamma_h )^{3/2} \Big) \nonumber\\
&+& 6 \Big(\sum_{h=1}^{n-1}\,\,\sum_{h'=1}^{n-h-1} (1-\frac{h+h'}{n})\E(X_1X_{1+h}X_{1+h+h'}) \Big) \big/ \Big( \sqrt{n}( \gamma_0+2\sum_{h=1}^n\big(1-\frac{h}{n}\big)\gamma_h )^{3/2}\Big).\nonumber\\
\end{eqnarray}
By virtue of the preceding representation of the skewness of  $T_{n,X}$, one can readily see that     $\beta_{T_{n,X}} \to 0$, as $n \to \infty$. This  is   in  agreement with the fact that, under appropriate conditions,  $T_{n,X}$ has a normal limiting distribution.
\\
We now introduce a randomization approach to construct  more symmetrical, and hence more accurate (cf. Theorem \ref{Edgeworth Expansion}),  versions of  $T_{n,X}$, based on the data set $X_1,\ldots,X_n$, as in (\ref{eq 1}),   as follows:

\begin{equation}\label{randomT}
T_{n,X,w}(\theta_n)=\big(\sum_{i=1}^n(w_i -\theta_n)(X_i-\mu)\big) \big/ \sqrt{ n \mathfrak{D}_{n,X,w}},
\end{equation}
where,

\begin{eqnarray}
\mathfrak{D}_{n,X,w} &:=&  \E(w_1 -\theta_n)^2\gamma_0+2  \E \big( (w_1 -\theta_n)(w_2 -\theta_n) \big)
\sum_{h=1}^n (1-\frac{h}{n})\gamma_h  \nonumber\\
&=& \V ( \sum_{i=1}^n (w_{i}-\theta_n) X_i)/n.
\end{eqnarray}
\\
The real valued constant $\theta_n$,   to which we shall refer as the window constant, and the random weights used in the definition of the randomized pivot $T_{n,X,w}(\theta_n)$, as in (\ref{randomT}),  are to be determined in view of the following scenario, to which we shall refer as the scheme {\textsf{(\textbf{RS})}}.

\subsection{The Randomization Scheme} \label{The Method*}
{\textsf{(\textbf{RS})}} ~~~~~~ Let $w_1,\ldots,w_n$ be either   nonsymmetric and  i.i.d. random variables with  $\E\vert w_1\vert^3<\infty$ or have a symmetric multinomial distribution, i.e.,  $\mathcal{M}ultinomial(n; 1/n, \ldots, 1/n)$.  Furthermore, we assume $w_{i}$ are \emph{independent} from the data $X_i$, $1\leq i \leq n$. Define

\begin{equation}\label{the factor}
\mathcal{H}_{n,X,w} (\theta):=
 \frac{\E \Big( \sum_{i=1}^n (w_{i}-\theta) X_i  \Big)^3}{n}. \nonumber
\end{equation}
Choose the window constant $\theta_n$,    such that $\mathcal{H}_{n,X,w} (\theta_n)= 0$ or   $\mathcal{H}_{n,X,w} (\theta_n) = \delta_n$, where $\delta_n \rightarrow 0 $, as $n \rightarrow \infty$.

\noindent
The following relation (\ref{Tw}) explains how  the randomization scheme {\textsf{(\textbf{RS})}}   results in more symmetrical pivotal quantities  $T_{n,X,w}(\theta_n)$, as in (\ref{randomT}), as compared to their  non-randomized counterpart  $T_{n,X}$, as in (\ref{T_{n,X}}). To do so, we compute the skewness of $T_{n,X,w}(\theta_n)$. In what follows, for simplicity and without loss of generality,  we assume that $\mu=0$.

\begin{eqnarray}\label{Tw}
\beta_{T_{n,X,w}} (\theta_n)&:=&
 \E \Big( \sum_{i=1}^n (w_{i}-\theta_n) X_i  \Big)^3 \big/ \Big( \V_{X,w}( \sum_{i=1}^n (w_{i}-\theta_n) X_i)\Big)^{3/2}\nonumber\\
&=& \mathcal{H}_{n,X,w} (\theta_n)  \frac{n}{\Big( \V_{X,w}( \sum_{i=1}^n (w_{i}-\theta_n) X_i)\Big)^{3/2}}.
\end{eqnarray}
%
From the preceding   equation, it is evident that  choosing a window constant $\theta_n$ in accordance   with the randomization scheme {\textsf{(\textbf{RS})}},  translates in  values for $\beta_{T_{n,X,w}}(\theta_n)$ which are zero or negligible. In other words, when a window constant $\theta_n$ is  chosen based on the   scheme {\textsf{(\textbf{RS})}},  the randomized pivot  ${T_{n,X,w}(\theta_n)}$, as in (\ref{randomT}), will be more symmetrical than its classical counterpart $T_{n,X}$ as in (\ref{T_{n,X}}). The effect of this symmetrization on the  accuracy of the CLT for $T_{n,X,w}(\theta_n)$, as in (\ref{randomT}),  is discussed in Section \ref{EE Section}.
\\
It is important to note that in the context of the scheme {\textsf{(\textbf{RS})}}, the skewness of the partial sum of a linear process can be made arbitral small  without changing its dependence structure. The change in the  dependence structure happens when the window constant equals the common mean of the weights.
This remark is formulated in the following Corollary \ref{Corollary 1}.

\begin{corollary}\label{Corollary 1}
Consider the randomization scheme {\textsf{(\textbf{RS})}}.
\\
(a) (Existence of solution) For large sample size  $n$, the equation $\mathcal{H}_{n,X,w} (\theta)=0$ has at least one real valued solution $\theta_n$.
\\
(b) (Dependence structure Preservation) Let $\theta_n$ be a real solution to $\mathcal{H}_{n,X,w} (\theta)=\delta_n$, where $\delta_n \rightarrow 0$. Then, we have  $\lim_{n \rightarrow \infty} \theta_n \neq  E(w_{1}) $.
\end{corollary}

\begin{remark}
From Corollary \ref{Corollary 1} it is implied that $\theta_n$ cannot be equal to $E(w_1)$ in the context of the scheme {\textsf{\textrm{(}\textbf{RS}\textrm{)}}}. This results
in two important and desirable properties. Firstly, it prevents change in the covariance structure of the data after being multiplied by the non-centered random wights. More precisely, in view of {\textsf{\textrm{(}\textbf{RS}\textrm{)}}}, if the original data $X_i$ have a short memory structure then so is their randomized version $(w_{i}-\theta_n)X_i$ and the same is also true for randomized long memory data. Secondly, non-centered weights yield randomized  confidence intervals, for the parameter of interest $\mu=\E(X_1)$, whose lengths vanish   as the sample size increases to $\infty$ (we refer to  Subsection \ref{confidence intervals} for further details in this regard).
\end{remark}


\section{The effect of scheme {\textsf{(\textbf{RS})}} on the error of the CLT for randomized linear processes}\label{EE Section}
We use the Edgeworth expansion  for a class  of linear processes  to illustrate the refinement that  results from our randomization approach {\textsf{(\textbf{RS})}}, as in Subsection \ref{The Method*}, which is primarily designed  to reduce $\beta_{T_{n},X}$, the skewness of the classical pivot $T_{n,X}$, as in (\ref{skewness4}). The choice of the Edgeworth expansion  is due to the fact that it provides an explicit and direct link between the skewness of a partial sum pivot and the   error admitted by its CLT.
\\
The following Theorem \ref{Edgeworth Expansion} gives an Edgeworth expansion for the sampling distribution of partial sums of the random variables  $(w_{i}-\theta_n)(X_{i}-\mu)$, $1\leq i \leq n$, where  $X_1,\ldots,X_n$ are data from some short memory linear processes as in (\ref{eq 1}).

\begin{theorem} \label{Edgeworth Expansion}
Assume that the data $X_1, \ldots, X_n$ are as in (\ref{eq 1}). Let the weights $w_1,w_2,\cdots$,  be i.i.d. random variables, that are  independent from the data.
Also, let $\{\theta_n\}$  be a converging sequence of constants such that, $\lim_{n \rightarrow \infty}\theta_n \neq \E (w_1)$. Furthermore, assume that
\\ \\
$(i)~\textrm{either }\sum_{k=0}^\infty a_k (w_{k+1}-\theta_n) \neq 0,\, a.s.\, $ for all $n\geq 1$, and $\underset{u\to\infty}\limsup\left\vert\E\left(e^{iu\zeta_1}\right)\right\vert<1$,
\\

 or
\\

$X_1\neq0$ a.s., and $\underset{u\to\infty}\limsup\left\vert\E\left(e^{iu w_1}\right)\right\vert<1$,
\\ \\
$
(ii)~ There\ exists\ \ell>0\ such~ that\ \vert a_k\vert\le (1/\ell)e^{- \ell k},\qquad\textrm{for all }k=0,1,\ldots,
$
\\ \\
$
(iii)~\E|\zeta_1|^5<\infty,
$
\\ \\
$(iv)~ \E|w_1|^5<\infty$.
\\
\\
Then,  we have for all $x \in \mathbb{R}$,

\begin{equation}\label{EE}
P\left({T_{n,X,w}(\theta_n)}
\le x\right)-\Phi(x)= \beta_{T_{n,X,w}}(\theta_n) H(x)/\sqrt{n} + O(1/n),
\end{equation}
where  $T_{n,X,w}(\theta)$ and $\beta_{T_{n,X,w}}(\theta)$ are, respectively,  as in  (\ref{randomT}) and  (\ref{Tw}),  $\Phi$ is the distribution function  of the standard normal distribution and $H$ is a polynomial of degree 3.
\end{theorem}
\noindent
We note that the conditions $\underset{u\to\infty}\limsup\left\vert\E\left(e^{iu w_1}\right)\right\vert<1$ and $(iv)$ of Theorem \ref{Edgeworth Expansion},   pose  no practical restriction on the choice of the random weights, as the class of random weights that satisfy these conditions is virtually unlimited.
\\
\\
In view of  the randomization scheme {\textsf{(\textbf{RS})}}, as  in Subsection \ref{The Method*},  and the relation (\ref{EE}), the smaller the $\beta_{T_{n,X,w}}(\theta_n)$ is, the closer the error of the CLT of $T_{n,X,w}(\theta_n)$  will be to $O(1/n)$. In the case  the weights $w_i$ and the window constant  $\theta_n$ are chosen in such a way  that $\beta_{T_{n,X,w}}(\theta_n)=0$
then,  in view of (\ref{EE}), the CLT for $T_{n,X,w}(\theta_n)$ will be  second order accurate, i.e., of magnitude  $O(1/n)$.  On the other hand,  when the linear process
$X_t$, as in (\ref{eq 1}), is so that its  innovations satisfy the Cramer condition, i.e.,  $\underset{u\to\infty}\limsup\left\vert\E\left(e^{iu\zeta_1}\right)\right\vert<1$, and also it satisfies
 conditions $(ii)$ - $(iii)$ of Theorem \ref{Edgeworth Expansion} as well as  the condition

\begin{equation}\nonumber
(i^\prime)~ \sum_{k=0}^\infty a_k\neq 0,  \qquad \qquad \qquad \qquad \qquad \qquad \qquad \qquad \qquad \qquad \qquad \qquad \qquad \qquad \qquad \qquad \qquad \qquad \qquad \qquad \qquad \qquad \qquad
\end{equation}
then,   Theorem (2.8)  of G\"{o}tze and Hipp \cite{Gotze and Hipp} and its Corollary (2.9), in the univariate case,  imply that, for all $x\in \mathbb{R}$, the sampling distribution of   $T_{n,X}$, as in (\ref{T_{n,X}}),  admits the following Edgeworth expansion:

\begin{equation}\label{EEC}
P\left(T_{n,X} \le x\right)-\Phi(x)=\beta_{T_{n,X}} \ H(x)/\sqrt{n}+O(1/n),
\end{equation}
where $\beta_{T_{n,X}}$, as in (\ref{skewness4}), is the skewness of $T_{n,X}$.
\\
To compare the  Edgeworth expansion (\ref{EE}) to (\ref{EEC}), we consider linear processes whose innovations $\zeta_j$ satisfy the Cramer condition and $\sum_{k=0}^{\infty} a_k \neq 0$.  We also consider  weights that are non-degenerate  i.i.d. and continuous with a finite fifth  moment. In this case the first part of condition $(i)$ in Theorem \ref{Edgeworth Expansion} holds true. Now under the conditions  $(ii)$, $(iii)$, and $(iv)$, one can see that (\ref{EE}) yields a smaller error when  $\beta_{T_{n,X,w}}(\theta_n)$ is small. In particular,  when $\beta_{T_{n,X,w}}(\theta_n)=0$, then the randomized pivot $T_{n,X,w}(\theta_n)$ is accurate of the second order  while the classical  $T_{n,X}$  remains accurate only of the first order, i.e., of order $O(1/\sqrt{n})$.

\begin{remark}\label{Advantage 1}
We note that in view of Theorem \ref{Edgeworth Expansion}, a randomly weighted  partial sum of a linear process can admit an Edgeworth expansion even when the non-randomized partial sum of the original data fails to do so. This   is due to the fact that in order to have the expansion (\ref{EE}), the crucial Cramer condition for  the innovations $\zeta_j$, which is assumed in  G\"{o}tze and Hipp \cite{Gotze and Hipp}, can now be shifted  to the weights $w_i$.
In other words,   when the innovations of the  linear process at hand fail to satisfy the Cramer condition,  then the expansion (\ref{EEC}) may no longer be valid. On the other hand, for the same linear process, using some appropriate non-centered randomizing weights, as specified in Theorem \ref{Edgeworth Expansion},   yields  a valid Edgeworth expansion as in (\ref{EE}) for the resulting normalized randomized partial sum $T_{n,X,w}(\theta_n)$, provided that the original linear process is so that $X_1\neq 0$ a.s..
This     constitutes an    achievement in significantly  enlarging the class of linear processes for which  Edgewrorth expansion holds.
\end{remark}

\begin{remark}\label{Advantage 2}
Further to the advantage of randomization discussed in Remark \ref{Advantage 1},
it is also  worth noting that when constructed using some continuous random  weights, so that they satisfy the conventional  conditions  (\textit{iii}) and (\textit{iv}) of Theorem \ref{Edgeworth Expansion}, the randomized pivot $T_{n,X,w}(\theta)$  admits     the Edgeworth expansion (\ref{EE})  for the linear processes   (\ref{eq 1}) even when   $\sum_{k=0}^\infty a_k=0$, provided that   (\textit{ii})  holds true. This is true, since, in the case that the random weights are continuous, the first part of condition  (\textit{i}) of Theorem \ref{Edgeworth Expansion} holds true even when $\sum_{k=0}^\infty a_k=0$. In contrast,   the Edgewroth expansion (\ref{EEC}) for the  classical  $T_{n,X}$, as in (\ref{T_{n,X}}), is valid only for short memory linear processes (\ref{eq 1}) for    which we have $\sum_{k=0}^\infty a_k \neq 0$, provided that the innovation satisfy the Cramer condition and, conditions  (\textit{ii}) and  (\textit{iii}) of Theorem \ref{Edgeworth Expansion} also hold true.
\end{remark}
\noindent
We conclude this section by noting that in the context of Theorem \ref{Edgeworth Expansion} and under its conditions $(i)$ and $(ii)$,  when $\E|\zeta_1|^4<\infty$ and $\E|w_1|^4< \infty$, $T_{n,X,w}(\theta)$, as in (\ref{randomT}), for all $x\in \mathds{R}$, admits the    expansion

\begin{equation}\label{newEE}
P\left({T_{n,X,w}(\theta_n)}
\le x\right)-\Phi(x)= \beta_{T_{n,X,w}}(\theta_n) H(x)/\sqrt{n} + o(1/\sqrt{n}).
\end{equation}
The preceding expansion is a counterpart of (\ref{EE}) on assuming the existence    of a finite fourth  moment for the innovations and the weights. The proof of the validity of (\ref{EE}) also results  from the proof of  Theorem \ref{Edgeworth Expansion} given in Section \ref{proofs}.
\\
The discussions following Theorem \ref{Edgeworth Expansion} on  the error reduction effect of the   scheme {\textsf{(\textbf{RS})}}, remain valid in view of (\ref{newEE})  with $o(1/\sqrt{n})$ instead of $O(1/n)$. Moreover,  the advantages of the randomization discussed in
Remarks  \ref{Advantage 1} and \ref{Advantage 2}, also remain valid when $T_{n,X,w}(\theta)$ admits the expansion (\ref{newEE}).

\section{CLT for randomized linear processes}\label{The CLT}
In this section we establish a CLT for the randomized pivot  $T_{n,X,w}(\theta_n)$, as in (\ref{randomT}), under less stringent conditions than those assumed in
Theorem \ref{Edgeworth Expansion}. The CLT in this section is valid for the random weights and window constants  as  characterized in  the randomization scheme {\textsf{(\textbf{RS})}} in Subsection \ref{The Method*}.
\\
For the use in   Theorems \ref{CLT},  \ref{CLT G^*^stu} and  \ref{Complete Studentization thm},
we let the notation $(\Omega_w,\mathcal{F}_w,P_w)$ stand for the  probability space on which the   triangular or non-triangular random weights  are defined.

\begin{theorem}\label{CLT}
Let $X_1,\ldots, X_n$ be the first $n$ terms of the linear process (\ref{eq 1})
 and consider the randomized pivot  $T_{n,X,w}(\theta_n)$, as in (\ref{randomT}), with the    weights $w_1,\ldots,w_n$  and the window constant $\theta_n$ be as specified in the
scheme {\textsf{(\textbf{RS})}}.
Then,   as $n\to\infty$, we have for all $x \in \mathbb{R}$
\begin{equation}\label{CLT Conditional}
P\big( T_{n,X,w}(\theta_n) \leq x~ \big| w_1,\ldots,w_n  \big) \longrightarrow \Phi(x) \qquad in\ probability-P_w
\end{equation}
and, consequently,
\begin{equation}\label{CLT Unconditional}
P\big( T_{n,X,w}(\theta_n) \leq x  \big) \longrightarrow \Phi(x).
\end{equation}
\end{theorem}
\noindent


\subsection{Studentization}
We introduce the following $G_{n,X,w}(\theta_n,d)$ as a  Studentized version of $T_{n,X,w}(\theta_n)$ that is valid for short and long memory processes as specified in Theorem \ref{CLT G^*^stu}.

\begin{equation}\label{Studentized}
G_{n,X,w}(\theta_n,d)=n^{-1/2-d} \Big(\sum_{i=1}^n(w_i-\theta_n)(X_i-\mu) \Big)  \big/ \sqrt{q^{-2d} \ \mathcal{S}_{n,q,w}   },
\end{equation}
where $q=O(n^{1/2})$,
\begin{equation}\nonumber
\mathcal{S}_{n,q,w}=\E (w_1-\theta_n)^2\bar\gamma_0+ \E\big( (w_1-\theta_n)(w_2-\theta_n) \big)\sum_{h=1}
^q\bar\gamma_h(1-\frac{h}{q})
\end{equation}
and $\bar{\gamma}_s:=\sum_{j=1}^{n-s}(X_j -\bar{X}_{n})(X_{j+s}-\bar{X}_n)\big/ n, \ 0\leq s \leq n-1, $ are the sample autocovariances.
\\
The Studentized pivot $G_{n,X,w}(\theta_n,d)$ can also  be viewed as a randomized version of the    Studentized pivotal quantity

\begin{equation}\label{classic Studentized}
T^{stu}_{n} (d):= n^{1/2 -d} (  \bar{X}_n -\mu) \Big/ \big( q^{-2d}\big( \bar{\gamma}_0  + 2 \sum_{h=1}^{q} \bar{\gamma}_{h}  (1-h/q)\big)  \big)^{1/2}.
\end{equation}

\noindent
Our   Studentizing sequence $q^{-2d} \mathcal{S}_{n,q,w}$ for $G_{n,X,w}(\theta_n,d)$ is a generalization
of $q^{-2d} \big(\bar{\gamma}_0  + 2  \sum_{h=1}^{q} \bar{\gamma}_{h}  (1-h/q)\big)$, the Studentzing sequence  for the non-randomized  $T^{stu}_{n} (d)$. Introduced in Abadir \textit{et al.} \cite{HACMAC}, the random sequence $ q^{-2d}\big( \bar{\gamma}_0  + 2  \sum_{h=1}^{q} \bar{\gamma}_{h}  (1-h/q)\big)$ is an   adaptation of   Bartlett-kernel heteroscedasticity and autocorrelation consistent (HAC) estimator (see, for example, Andrews  \cite{Andrews}) that allows for long memory.  Although, in this paper attention is restricted   to our HAC-type Studentizing sequences of the form $q^{-2d}\mathcal{S}_{n,q,w}$, we remark that it is also desirable to use other  existing robust sample based estimators  such as Robinson's periodogram based   memory and autocorrelation consistent (MAC) estimator (cf. Robinson \cite{Robinson}), to construct  Studentizing sequences for the randomized sum $n^{-1/2-d} \big(\sum_{i=1}^n(w_i-\theta_n)(X_i-\mu) \big)$.
\\
In what follows the notation $G_{n,X,w}(\theta_n,\hat d)$ in the case of a long memory  process will stand for $G_{n,X,w}(\theta_n,d)$, where the memory parameter $d$ is replaced by its sample based estimate  $\hat d$  as specified in the following Theorem \ref{CLT G^*^stu} which establishes  the CLT for  $G_{n,X,w}(\theta_n,d)$ and it reads as follows.

\begin{theorem}\label{CLT G^*^stu}
Consider  $X_1, \ldots, X_n$, the first $n$ terms of the linear process (\ref{eq 1}), and  let the weights $w_1,\cdots,w_n$ and the window constants $\theta_n$ be as in Theorem \ref{CLT}.
\\
\\
(A) If   the   linear process (\ref{eq 1}) is of  short memory, i.e., $\sum_{k=0}^{\infty} |a_{k}|<\infty$,  and   $\E \zeta_{1}^{4}<\infty$,
 then, as $n, q \to \infty$ such that $q=O(n^{1/2})$,    we have, for all $x \in \mathbb{R}$,
\begin{equation}\nonumber\label{eq 18}
P\big(G_{n,X,w}(\theta_n,0)\leq x ~ \big|w_1,\ldots,w_n\big) \longrightarrow \Phi(x) \ in \ probability-P_w,
\end{equation}
and, consequently,
\begin{equation}\nonumber\label{eq 19}
P(G_{n,X,w}(\theta_n,0)\leq x) \longrightarrow \Phi(x),\ t \in \mathbb{R}.
\end{equation}
(B) Let  the linear process (\ref{eq 1})    be of long memory  such that $\E \zeta_{1}^{4}<\infty$ and, as $k \rightarrow \infty$,  $a_k \sim c k^{d-1}$, for some $c>0$, where $0< d <1/2$. Then, as $n,q\to \infty$ such that $q=O(n^{1/2})$,  we have for all $x \in \mathbb{R}$,

\begin{eqnarray*}
&&P\big(G_{n,X,w}(\theta_n,d)\leq x~ \big| w_1,\ldots,w_n\big) \longrightarrow \Phi(x) \ in \ probability-P_w,\\
&&P\big(G_{n,X,w}(\theta_n,\hat{d})\leq x ~ \big| w_1,\ldots,w_n\big) \longrightarrow \Phi(x) \ in \ probability-P_w,
\end{eqnarray*}
and, consequently,

\begin{eqnarray*}
&&P(G_{n,X,w}(\theta_n,d)\leq x) \longrightarrow \Phi(x),\ x \in \mathds{R}, \\
&&P(G_{n,X,w}(\theta_n,\hat{d})\leq x) \longrightarrow \Phi(x), \ x \in \mathds{R},
\end{eqnarray*}
where $\hat{d}$ is an estimator of the memory parameter $d$ such that $\hat{d}-d=o_{P}(1/\log n)$.
\end{theorem}
\noindent
When the linear process in Theorem \ref{CLT G^*^stu} is of long memory with memory parameter $d$,  there are a number of  estimators $\hat{d}$ in the literature that can be used to estimate $d$. The MLE  of $d$, with the Haslett and Raftery  \cite{Haslett and Raftery} method used to approximate the likelihood and, Whittle estimator (cf. K\"{u}nsch \cite{Kunch1987} and Robinson  \cite{Robinson dhat}) are two examples of the estimators of the memory parameter $d$. For more on estimators
for the memory parameter  and their asymptotic behavior, we refer to Bhansali and Kokoszka  \cite{Bhansali and Kokoszka},    Moulines and Soulier \cite{Moulines and Soulier} and references therein.

\subsection{Randomized confidence intervals}\label{confidence intervals}
By virtue of Theorem \ref{CLT G^*^stu}, for the parameter of interest $\mu$, the mean of the  linear process (\ref{eq 1}), the Studentized  randomized pivotal quantity $G_{n,X,w}(\theta_n,d)$, as in (\ref{Studentized}),  yields asymptotically exact size randomized confidence intervals with nominal size $100(1-\alpha)\%$, $0<\alpha<1$,  which are  of the form:

\begin{equation}\label{randomized CI}
I_{n,X,w}(\theta_n) := \Big[ \min\{\mathcal{A}_{n,X,w,\alpha}(\theta_n), \mathcal{B}_{n,X,w,\alpha}(\theta_n) \} , \max\{\mathcal{A}_{n,X,w,\alpha}(\theta_n), \mathcal{B}_{n,X,w,\alpha}(\theta_n)\}   \Big],
\end{equation}
where

\begin{eqnarray*}
\mathcal{A}_{n,X,w,\alpha}(\theta_n) &=& \Big(\sum_{i=1}^n (w_{i}-\theta_n)X_i -z_{\alpha/2} n^{1/2+d} \sqrt{q^{-2d} \ \mathcal{S}_{n,q,w}   }   \Big) \big/ \Big(\sum_{i=1}^n ( w_{i}  -\theta_n ) \Big)\\
\mathcal{B}_{n,X,w,\alpha}(\theta_n) &=& \Big(\sum_{i=1}^n (w_{i}-\theta_n)X_i +z_{\alpha/2} n^{1/2+d} \sqrt{q^{-2d} \ \mathcal{S}_{n,q,w}   }   \Big) \big/ \Big(\sum_{i=1}^n ( w_{i}  -\theta_n ) \Big),
\end{eqnarray*}
and $z_{\alpha/2}$ is the $100(1-\alpha/2)$-th percentile of the standard normal distribution.
The length of the randomized confidence interval $I_{n,X,w}(\theta_n)$ is

\begin{equation}\label{length randomized CI}
length(I_{n,X,w}(\theta_n)) = 2 z_{\alpha/2} n^{1/2+d} \sqrt{q^{-2d}  \mathcal{S}_{n,q,w}   }    \big/ \big| \sum_{i=1}^n ( w_{i}  -\theta_n ) \big|.
\end{equation}

\begin{remark}\label{on the length}
Along the lines of the proof of  Theorem \ref{CLT G^*^stu} in Section \ref{proofs} it is shown that
for $0 \leq d <1/2$,  $q^{-2d}  \mathcal{S}_{n,q,w}$ asymptotically coincides with $\textrm{Var}\big(n^{-1/2-d} \sum_{i=1}^n (w_{i}-\theta_n) X_i\big)$. In view of the latter asymptotic  equivalence together with  (\ref{length randomized CI}),
one can  see that when $\theta_n=\E(w_{1})$, the length of the randomized confidence interval $I_{n,X,w}(\theta_n)$ fails to  vanish as $n,q \to \infty$ such that $q=O(n^{1/2})$. In view of Corollary \ref{Corollary 1}, choosing $\theta_n$ to be equal to $\E(w_{1})$ is not possible in the framework of the scheme {\textsf{(\textbf{RS})}}, as in Subsection \ref{The Method*}. Therefore, {\textsf{(\textbf{RS})}}  results in randomized confidence intervals whose limiting lengths  is zero as $n,q \to \infty$ such that $q=O(n^{1/2})$.
\end{remark}
\noindent
Tables 1  and 2 below provide an empirical comparison of   the  performance of the randomized confidence interval $I_{n,X,w}(\theta_n)$,  as in (\ref{randomized CI}), and that of

\begin{equation}\label{classic CI}
I_{n,X}=\bar{X}_n \pm z_{\alpha/2} n^{-1/2+d} \sqrt{q^{-2d}\big( \bar{\gamma}_0  + 2 \sum_{h=1}^{q} \bar{\gamma}_{h}  (1-h/q)\big)},
\end{equation}
which is  constructed based on the classical $T^{stu}_{n} (d)$, as in (\ref{classic Studentized}). The lag-length or the bandwidth  $q$ was chosen  based on relation (2.14) of  Abadir \emph{et al}. \cite{HACMAC} in each case. More precisely,  in Tables 1 and 2 we let  $q$ be $ceiling(n^{1/3})$ for the examined  short memory linear process.
\\
It is easy to see that the length of   $I_{n,X}$ has the form:

\begin{equation}\label{length classic CI}
length(I_{n,X})= 2 z_{ \alpha/2} n^{-1/2+d} \sqrt{q^{-2d}\big( \bar{\gamma}_0  + 2 \sum_{h=1}^{q} \bar{\gamma}_{h}  (1-h/q)\big)}.
\end{equation}
The results in Tables 1 and 2   are   at the nominal level of \verb|95|\% with $\verb|z|_{\alpha/2}=\verb|1.96|$, and     based on  2000  replications of the therein specified short memory \verb|AR(1)| data, i.e., when  $X_t=\phi X_{t-1}+\zeta_t$, where $\zeta_t$ have  standardized, heavily skewed,  \verb|lognormal(0,1)| distribution. For the randomized confidence intervals in Tables 1 and 2, 2000 sets of therein specified   random weights also   generated simultaneously with the data.
We note in passing that the   processes considered   in Tables 1 and 2 are of the form (\ref{eq 1}).

\begin{table}[h!]\label{Table 1}
\vspace{- .2 cm}
\begin{center}
\begin{tabular}{||l|| l|| ccc}
\multicolumn{1}{c}{} &  \multicolumn{1}{c}{} &$\verb|I|_{n,X,w}(\theta_n)$ &   & $\verb|I|_{n,X}$   \\ \cline{1-5}
              &\cellcolor[gray]{0.8}\verb|weights|   &\cellcolor[gray]{0.8}\verb|Bernoulli(1/4)| & \cellcolor[gray]{0.8}  &\cellcolor[gray]{0.8}\verb|NA| \\
              &   $\theta_n$            & \verb|1/4+0.14| &  &\verb|NA| \\
  \verb|n=200|
              &         \verb|length|        & \verb|1.81| &  &\verb|1.43| \\
              &\cellcolor[gray]{0.8}\verb|coverage|      & \cellcolor[gray]{0.8} \verb|0.9155| &\cellcolor[gray]{0.8}  &\cellcolor[gray]{0.8} \verb|0.8615|\\
   \hline\hline

             &  \cellcolor[gray]{0.8}\verb|weights|   & \cellcolor[gray]{0.8}\verb|Bernoulli(1/4)| &  \cellcolor[gray]{0.8}& \cellcolor[gray]{0.8}\verb|NA| \\

              &   $\theta_n$            & \verb|1/4+0.1| &  &\verb|NA| \\
  \verb|n=400|

              &         \verb|length|  & \verb|1.68| &  &\verb|1.11| \\

              &\cellcolor[gray]{0.8}\verb|coverage| &\cellcolor[gray]{0.8}\verb|0.9365| & \cellcolor[gray]{0.8}  & \cellcolor[gray]{0.8}\verb|0.8865|\\
   \cline{1-5}
\end{tabular}
\end{center}
\vspace{-.2 cm}\caption{ AR(1) with $\phi=0.8$}
 \end{table}


\begin{table}[h!]\label{Table 3}
\begin{center}
\begin{tabular}{||l|| l|| ccc}
\multicolumn{1}{c}{} & \multicolumn{1}{c}{} &$\verb|I|_{n,X,w}(\theta_n)$ &   & $\verb|I|_{n,X}$   \\ \cline{1-5}
              &\cellcolor[gray]{0.8}\verb|weights|   &\cellcolor[gray]{0.8}\verb|Multinomial(n;1/n,...,1/n)| & \cellcolor[gray]{0.8}  &\cellcolor[gray]{0.8}\verb|NA| \\
              &   $\theta_n$            & \verb|1-0.27| &  &\verb|NA| \\
  \verb|n=200|
              &         \verb|length|        & \verb|1.91| &  &\verb|1.43| \\
              &\cellcolor[gray]{0.8}\verb|coverage|      & \cellcolor[gray]{0.8} \verb|0.912| &\cellcolor[gray]{0.8}  &\cellcolor[gray]{0.8} \verb|0.866|\\
   \hline\hline
             &  \cellcolor[gray]{0.8}\verb|weights|   & \cellcolor[gray]{0.8}\verb|Multinomial(n;1/n,...,1/n)| &  \cellcolor[gray]{0.8}& \cellcolor[gray]{0.8}\verb|NA| \\
              &   $\theta_n$            & \verb|1+0.23| &  &\verb|NA| \\
  \verb|n=400|
              &         \verb|length|        & \verb|1.59| &  &\verb|1.11| \\
              &\cellcolor[gray]{0.8}\verb|coverage| &\cellcolor[gray]{0.8}\verb|0.942| & \cellcolor[gray]{0.8}  & \cellcolor[gray]{0.8}\verb|0.8995|\\
   \cline{1-5}
\end{tabular}
\end{center}
\vspace{-.2 cm}\caption{AR(1) with $\phi=0.8$}
 \end{table}


\section{Comparison to bootstrap confidence intervals  for long memory linear processes}\label{Comparison to bootstrap confidence intervals  for long memory linear processes}
In the context of long memory data, the residual resampling method, known as the sieve bootstrap (cf., e. g., Poskitt  \cite{Pos2008}), is built around approximating a  long memory process  by approximating its infinite auto regressive representation by  a truncated one. However, this approximation  produces error terms which tend to be significant (cf. Table 3).  The approximation significantly underestimates the variance of the statistic of interest, which is the sample mean in the context of this paper, based on long memory linear process structured data. The underestimation of the variance translates in poor coverage probabilities of the confidence intervals constructed using the   (raw) sieve, as can be seen in Table 3.   Filtered sieve  introduced by Poskitt \emph{et al.}   \cite{Pos} is a modification of the sieve that tends to improve upon the coverage probability of the  sieve. It essentially  consists of applying the raw sieve to a filtered series (basically to a truncated version of the infinite sum $(1-B)^{\hat d} X_t$, where $\hat d$ is an estimator of the long memory parameter $d$) and then  unfilter the resulting  bootstrapped error series $\hat\epsilon_t^*$  (again by considering the truncated  sum $(1-B)^{-\hat d}\hat\epsilon_t^*$ ).
\\
The block bootstrap is another method of drawing a bootstrap sample from a given set of   dependent and temporal  observations  with the goal of preventing the i.i.d. nature of the naive bootstrap from  dismissing the dependence and chronological order of the data. The method   essentially  consists of first partitioning the data into a number of blocks and then ressampling from these blocks. This method was first introduced by K\"{u}nsch  \cite{Kunchblock} for short memory linear processes. The validity of this resampling scheme for both short and long memory linear processes has been investigated in a number of papers, e. g., G\"{o}tze and K\"{u}nsch \cite{Gotze and Kunch}  and Kim and Nordman \cite{Kim and Nordman}.
\\
Although, with a relatively large number of replications, the existing bootstrap methods tend to perform  well for short memory data, it is not the case when they are applied to long memory  data. In our numerical study below, we consider three methods of constructing bootstrap confidence intervals for the mean of a population from which  long memory structured linear process data sets are simulated. In Table 3  we compare  the performance of the resulting confidence intervals from the  three bootstrap methods    sieve, filtered or augmented sieve and the block bootstrap for long memory linear processes to those   of the randomized confidence interval  $I_{n,X,w}(.)$, as in (\ref{randomized CI}),  and the classical $I_{n,X}$, as in (\ref{classic CI}), in terms  of their respective empirical probabilities of coverage.
\\
In Table 3, $I_{AugSiv}$, $I_{Bloc}$ and $I_{Siv}$ stand, respectively, for the augmented sieve, block-bootstrap and the sieve confidence intervals at nominal level of 95\%. Each of these bootstrap confidence intervals is constructed based on $B=1000$ bootstrap replications   to estimate the cut-off points for each of the 2000 sets of simulated data. In Table 3, $\mathcal{M}ult$ is a short hand notation for the symmetric multinomial distribution, i.e., $\mathcal{M}ultinomial(n;1/n,\cdots,1/n)$. For all examined confidence intervals, the simulated observations in Table 3,  are generated from  the  fractionally integrated model   $\verb|X|_t =\verb|(1-B)|^{-\verb|d|} \zeta_t$, where, $\verb|B|$ is the back-shift operator, i.e., $\verb|B| \zeta_t= \zeta_{t-1}$,   \verb|d| is the memory parameter
and $\zeta_t$ have  standardized  \verb|lognormal(0,1)| distribution. Similarly to Tables 1 and 2, to construct the randomized confidence interval $I_{n,X,w} (\theta_n)$, for each   simulated set of data, simultaneously,  a set of random weights from symmetric  multinomial distribution was  generated. In Table 3 for the therein considered  long memory linear process  with  $d=0.4$, we let $q$ be    $ceiling(n^{1/2 -d})$.

\begin{table}[h!]\label{Table 3}
\begin{center}
\begin{tabular}{||l|| l|| cccccc}
\multicolumn{1}{c}{} & \multicolumn{1}{c}{} &$\verb|I|_{n,X,w}(\theta_n)$ &   & $\verb|I|_{n,X}$  & $I_{Bloc}$ & $I_{AugSiv}$ & $I_{Siv}$\\ \cline{1-8}
              &\cellcolor[gray]{0.8}\verb|weights|   &\cellcolor[gray]{0.8} $\mathcal{M}ult$
                & \cellcolor[gray]{0.8}  &\cellcolor[gray]{0.8}\verb|NA| & \cellcolor[gray]{0.8} \verb|NA| & \cellcolor[gray]{0.8} \verb|NA|
                 & \cellcolor[gray]{0.8} \verb|NA|\\
              &   $\theta_n$            & \verb|1+0.97| &  &\verb|NA|    &\verb|NA|  & \verb|NA| & \verb|NA| \\
  \verb|n=100|
              &         \verb|length|        & \verb|3.28| &  &\verb|2.50| & \verb|2.26| & \verb|1.90| & \verb|0.8695|\\
              &\cellcolor[gray]{0.8}\verb|coverage|      & \cellcolor[gray]{0.8} \verb|0.935| &\cellcolor[gray]{0.8}  &\cellcolor[gray]{0.8} \verb|0.8525| &  \cellcolor[gray]{0.8} \verb|0.803| & \cellcolor[gray]{0.8} \verb|0.710| & \cellcolor[gray]{0.8} \verb|0.3795| \\
   \hline\hline
             &  \cellcolor[gray]{0.8}\verb|weights|   & \cellcolor[gray]{0.8} $\mathcal{M}ult$ &  \cellcolor[gray]{0.8}& \cellcolor[gray]{0.8}\verb|NA|   & \cellcolor[gray]{0.8} \verb|NA| & \cellcolor[gray]{0.8} \verb|NA| & \cellcolor[gray]{0.8} \verb|NA|\\
              &   $\theta_n$            & \verb|1+0.97| &  &\verb|NA|   & \verb|NA| & \verb|NA| & \verb|NA|\\
  \verb|n=200|
              &         \verb|length|        & \verb|3.20| &  &\verb|2.45| & \verb|2.25| &  \verb|1.83| & \verb|0.8| \\
              &\cellcolor[gray]{0.8}\verb|coverage| &\cellcolor[gray]{0.8}\verb|0.9445| & \cellcolor[gray]{0.8}  & \cellcolor[gray]{0.8}\verb| 0.864|  & \cellcolor[gray]{0.8} \verb|0.823| & \cellcolor[gray]{0.8} \verb|0.739| & \cellcolor[gray]{0.8} \verb|0.4035| \\
   \cline{1-8}
\end{tabular}
\end{center}
\vspace{-.2 cm}\caption{ Fractionally integrated long memory with d=0.4}
\end{table}

\begin{remark}
As it is evident from the numerical study presented in Table 3, the randomized confidence interval $I_{n,X,w}(\theta)$, introduced in this paper, as in (\ref{randomized CI}),     produces significantly more accurate confidence intervals as compared to the classical $I_{n,X}$, as in (\ref{classic CI}), and the three examined bootstrap confidence intervals. It is   noteworthy that the higher accuracy of $I_{n,X,w}(\theta)$ can also be achieved using other random weights than the symmetric multinomial used in Table 3, provided that they are chosen according to the skewness reduction   scheme    {\textsf{(\textbf{RS})}}, as in Subsection  \ref{The Method*}.
\end{remark}

\section{Complete Studentization}\label{Complete Studentization}
In the Studentizing sequence $q^{-2d}\mathcal{S}_{n,q,w}$ used in the definition of the    Studentized randomized  pivot $G_{n,X,w}(\theta_n,d)$, as  defined  in (\ref{Studentized}), we used the actual values of the  moments of the wights, i.e., $\E(w_{1})$, $\E(w_{1})^2$ and  $\E( w_{1} w_{2})$. The motive of the   use of such a partial Studentization  is justified  considering that in  the framework created by the scheme    {\textsf{(\textbf{RS})}},  the distribution of the  weights are usually   known. Despite the validity of this reasoning, it is also desirable to   investigate the asymptotic behavior of the completely Studentized version of $G_{n,X,w}(\theta_n,d)$. In the following Theorem \ref{Complete Studentization thm} we establish a   CLT result for the completely Studentized randomized pivotal quantity

\begin{equation}\label{Completely Studentized}
\hat{G}_{n,X,w}(\theta_n,d)=n^{-1/2-d} \Big(\sum_{i=1}^n(w_i-\theta_n)(X_i-\mu) \Big)  \big/ \sqrt{q^{-2d} \ \hat{\mathcal{S}}_{n,q,w}   },
\end{equation}
where $q=O(n^{1/2})$,
\begin{equation}\nonumber
\hat{\mathcal{S}}_{n,q,w}=\frac{1}{n}\sum_{j=1}^{n} (w_j-\theta_n)^2\bar\gamma_0+ 2 q^{-1} \sum_{h=1}^{q} \bar{\gamma}_h  \sum_{j=1}^{q-h} (w_j-\theta_n)(w_{j+h}-\theta_n),
\end{equation}
where $\bar{\gamma}_s$, $ 0\leq s \leq n-1, $ are the sample autocovariances, as defined right after (\ref{Studentized}).

\begin{theorem}\label{Complete Studentization thm}
Consider  $X_1, \ldots, X_n$, the first $n$ terms of the linear process (\ref{eq 1}), and  let the weights $w_1,\cdots,w_n$ and the window constants $\theta_n$ be as in Theorem \ref{CLT}.
\\
\\
(A) If   the   linear process (\ref{eq 1}) is of  short memory, i.e., $\sum_{k=0}^{\infty} |a_{k}|<\infty$,  and   $\E \zeta_{1}^{4}<\infty$,
 then, as $n, q \to \infty$ such that $q=O(n^{1/2})$,    we have, for all $x \in \mathbb{R}$,
\begin{equation}\nonumber
P\big(\hat{G}_{n,X,w}(\theta_n,0)\leq x ~ \big|w_1,\ldots,w_n\big) \longrightarrow \Phi(x) \ in \ probability-P_w,
\end{equation}
and, consequently,
\begin{equation}\nonumber
P(\hat{G}_{n,X,w}(\theta_n,0)\leq x) \longrightarrow \Phi(x),\ t \in \mathbb{R}.
\end{equation}
(B) Let  the linear process (\ref{eq 1})    be of long memory  such that $\E \zeta_{1}^{4}<\infty$ and, as $k \rightarrow \infty$,  $a_k \sim c k^{d-1}$, for some $c>0$, where $0< d <1/2$. Then, as $n,q\to \infty$ such that $q=O(n^{1/2})$,  we have for all $x \in \mathbb{R}$,

\begin{eqnarray*}
&&P\big(\hat{G}_{n,X,w}(\theta_n,d)\leq x~ \big| w_1,\ldots,w_n\big) \longrightarrow \Phi(x) \ in \ probability-P_w,\\
&&P\big(\hat{G}_{n,X,w}(\theta_n,\hat{d})\leq x ~ \big| w_1,\ldots,w_n\big) \longrightarrow \Phi(x) \ in \ probability-P_w,
\end{eqnarray*}
and, consequently,

\begin{eqnarray*}
&&P(\hat{G}_{n,X,w}(\theta_n,d)\leq x) \longrightarrow \Phi(x),\ x \in \mathds{R}, \\
&&P(\hat{G}_{n,X,w}(\theta_n,\hat{d})\leq x) \longrightarrow \Phi(x), \ x \in \mathds{R},
\end{eqnarray*}
where $\hat{d}$ is an estimator of the memory parameter $d$ such that $\hat{d}-d=o_{P}(1/\log n)$.
\end{theorem}
\noindent
The  CLTs in the preceding Theorem \ref{Complete Studentization thm}  are counterparts of those in  Theorem \ref{CLT G^*^stu}.

\section{Proofs}\label{proofs}

\subsection*{Proof of  Corollary \ref{Corollary 1}}
The third degree polynomial, in $\theta$, $\mathcal{H}_{n,X,w}(\theta)$ can be written as:
\begin{eqnarray}\label{short1}
 \mathcal{H}_{n,X,w}(\theta)&=&\frac{1}{n}\mathbb{E} \big(\sum_{i=1}^n(w_i-\theta)X_i\big)^3
\nonumber\\
&=&-\frac{1}{n}\mathbb{E} \big(\sum_{i=1}^n X_i \big)^3   \theta^3 \nonumber\\
&+&\frac{3}{n}\mathbb{E} \Big(  \big(\sum_{i=1}^n w_i  X_i \big)
\big(\sum_{i=1}^n X_i \big)^2 \Big)\theta^2
-\frac{3}{n}\mathbb{E} \Big(\big(\sum_{i=1}^n w_i  X_i \big)^2
\big(\sum_{i=1}^n X_i \big)\Big)\theta   \nonumber\\
&+&\frac{1}{n}\mathbb{E} \big(\sum_{i=1}^n w_i X_i \big)^3.
\end{eqnarray}
From the preceding representation, it can be  seen that  $\mathcal{H}_{n,X,w}(\theta)$
will have at least one real solution if     for large $n$ we have


\begin{equation}\label{real}
\frac{1}{n}\mathbb{E} \big(\sum_{i=1}^n X_i \big)^3 \neq 0.
\end{equation}
In order to show that,   as $n\to\infty$, (\ref{real}) holds true, we first consider the case  when the process $X_t$ is of short memory, i.e.

$$
X_t=\sum_{j=0}^\infty a_j\epsilon_{t-j},
$$
$\epsilon_t$ being i.i.d. white noise and
$$
\sum_{j=0}^\infty \vert a_j\vert<\infty\textrm{ and }\sum_{j=0}^
\infty a_j \neq 0.
$$
In this case, on assuming that the innovations are nonsymmetric  such that $\mathbb{E}(\epsilon_1^3)\neq 0$, it is easy to see that, as $n\to\infty$,
$$
\frac{1}{n}\mathbb{E}\left[\left(\sum_{i=1}^nX_i\right)^3\right]\to\mathbb{E}(\epsilon_1^3)\left(\sum_{j=0}^
\infty a_j\right)^3 \neq 0.
$$
Now we consider the case when $X_t$ is of long memory, i.e.
$$
a_i\sim i^{(d-1)},\qquad\textrm{ as } i\to\infty.
$$
Without loss of generality, we   assume that $a_i>0$ for all $i$, as there could only be finitely many $i$ for which $a_i \leq 0$, and such $i$'s will result in only a finite sum.
Now, similarly to the formula (\ref{Tw}),  on   assuming that $\mathbb{E}(\epsilon_1^3)>0$, as $n\to\infty$, we have
\begin{eqnarray*}\lefteqn{
\frac{1}{n}\mathbb{E}\left[\left(\sum_{i=1}^nX_i\right)^3\right]=\mathbb{E}(X_1^3)+3\sum_{h=1}^n\left(1
-\frac{h}{n}\right)\mathbb{E}(X_1^2X_{1+h}+X_1X^2_{1+h})}\\
&&+\sum_{h=1}^{n-1}\sum_{h'=1}^{n-h-1}
\left(1-\frac{h+h'}{n}\right)\mathbb{E}\left(X_1X_{1+h}X_{1+h+h'}\right)\\
&\ge&\sum_{h=1}^n\left(1
-\frac{h}{n}\right)\mathbb{E}(X_1^2X_{1+h})=
\sum_{h=1}^n\left(1-\frac{h}{n}\right)\mathbb{E}(\epsilon_1^3)\sum_{i=0}^\infty
a_i^2a_{i+h}\\
&\ge& a_{0}^2  \mathbb{E}(\epsilon_1^3)\sum_{h=1}^n\left(1-\frac{h}{n}\right)a_h \to\infty.
\end{eqnarray*}
In summary, we have proved  that for both short and long memory, for $n$ large, (\ref{real}) holds true. This completes the proof of part (a) of this corollary.
\\
Now if we take the weights $w_i $ to be i.i.d. and  nonsymmetric,  with   $\mathbb{E}\big(w_i -\mathbb{E}(w_i )\big)^3 \neq0$, we can see from (\ref{Tw}) that
$$
\mathcal{H}_{n,X,w}(\mathbb{E}(w_1 ))=\mathbb{E}\big(w_i -\mathbb{E}(w_i ) \big)^3 \mathbb{E}(X_1^3) \neq 0
$$
and hence $\mathbb{E}(w_1 )$ cannot be a root of the polynomial $\mathcal{H}_{n,X,w}(\theta)$.
\\
So, we conclude  that there is a real root  $\theta_n$ for $\mathcal{H}_{n,X,w}(\theta)$ and that this root is different from $\mathbb{E}(w_1 )$. Choosing this root as a value for $\theta$ will guarantee a second order accuracy of the Theorem \ref{Edgeworth Expansion}. Similar argument can be made when choosing the weights have a symmetric multinomial  distribution. This is true  since  in that case,    using  the joint moment generating function of $(w_1,\ldots,w_n$) and  the fact that $\mathbb{E}(w_i)=1$,     as $n\to\infty$, we have
$$
\mathcal{H}_{n,X,w}(\mathbb{E}(w_1))\to\mathbb{E}(X_1^3)\neq0,
$$
and therefore for $n$ large,  $\mathcal{H}_{n,X,w}(\mathbb{E}(w_1 )) \neq 0$, which again guarantees  that the real root we are seeking to achieve a negligible skewness, in the context of scheme {\textsf{(\textbf{RS})}},   cannot be equal to $\mathbb{E}(w_1)$. This completes the proof of part (b). Now the proof of the corollary is complete. $\square$

\subsection*{Proof of Theorem \ref{Edgeworth Expansion}}
We first define the following $\sigma$-fields to be used  in the proof of this theorem.

\begin{eqnarray*}
\mathcal{D}_j&:=&\sigma(w_{j-1},\zeta_j),\quad j=0,\pm1,\pm,2,\ldots,\\
\mathcal{D}_{j,n}&:=&\sigma(\zeta_n,w_{j-1},\zeta_j),\quad j=0,\pm1,\pm,2,\ldots,\\
\mathcal{D}_{p}^{q}&:=&\sigma\{\big( w_{j-1},\zeta_j  \big),\quad p\leq j\leq q\}.
\end{eqnarray*}
In order for the definition of $\mathcal{D}_j$ to hold true, we extended the weights $w_1,w_2,\ldots,$ to
$$\ldots,w_{-1},w_0,w_1,\ldots.$$
Theorem \ref{Edgeworth Expansion}   results  from  Theorem (2.8), and its Corollary (2.9), of G\"{o}tze and Hipp \cite{Gotze and Hipp}. More precisely, we show that under the conditions of Theorem \ref{Edgeworth Expansion},  conditions (2.3) - (2.6)  of G\"{o}tze and Hipp \cite{Gotze and Hipp}
 hold true. In our case, (2.4) of G\"{o}tze and Hipp \cite{Gotze and Hipp} holds true trivially as $\mathcal{D}_{-\infty}^{n}$ is independent from $\mathcal{D}_{n+m}^{\infty}$, for all $m\geq 1$. To see why (2.3) holds, observe that
\begin{eqnarray}\label{EEw}
\E \vert(w_n-\theta_n)\Big(X_n-\sum_{k=0}^m  a_k \zeta_{n-k}\Big)\vert
&\le& \E\vert w_1-\theta_n \vert \  \E \vert X_n-\sum_{k=0}^m a_k \zeta_{n-k}\vert \nonumber\\
&\le& \E\vert w_1-\theta_n \vert\mathbb{E}\vert\xi_1\vert (1/\ell)  e^{-\ell m}\le (c_n/\ell)  e^{-\ell m},
\end{eqnarray}
where $c_n=\E|w_1-\theta_n| \E|\zeta_1|$. If $c_n \le 1$, then (2.3) of G\"{o}tze and Hipp \cite{Gotze and Hipp} clearly holds. If $c_n>1$, then $c_n$ will be bounded above by some constant $c>0$, therefore  the right hand side of (\ref{EEw}) is bounded above by $(c/\ell) e^{-\ell m/c}$. This means that (2.3) of G\"{o}tze and Hipp \cite{Gotze and Hipp} still holds but with $\ell/c $ instead of $\ell$. In conclusion, condition (2.3) of G\"{o}tze and Hipp \cite{Gotze and Hipp} holds true in our context of Theorem \ref{Edgeworth Expansion}.
\\
We now  show that (2.5) of G\"{o}tze and Hipp \cite{Gotze and Hipp} also holds true in the context of Theorem \ref{Edgeworth Expansion}, noting first that in what follows $\E(.~/\mathcal{F})$ stands for conditional expected value given the $\sigma$-field $\mathcal{F}$.  Condition  (2.5) of G\"{o}tze and Hipp in our context corresponds to the following statement: For any $b>0$, there exists $\ell>0$ such that for $\vert u\vert>b$ and all $m,n$ such that $1/b < m <n$,  we have

\begin{eqnarray}\label{new2.5}
&&\E\Big\vert\E\Big(\exp\Big(iu\Big[(w_{n-m}-\theta_n)X_{n-m}+\cdots+(w_{n-1}-\theta_n)X_{n-1}\nonumber\\
&&~ ~ ~ ~ +(w_n-\theta_n)X_n+\cdots+(w_{n+m}-\theta_n) X_{n+m}\Big]\Big)\Big/\mathcal{D}_j,j\neq n\Big)\Big\vert\le e^{-\ell}.
\end{eqnarray}
To show that (\ref{new2.5}) holds true under the assumptions of our Theorem \ref{Edgeworth Expansion}, we let $\Psi_\zeta$ and $\Psi_w$  respectively denote the characteristic functions of $\zeta_n$ and $w_n$.
Observe now that the left hand side  of (\ref{new2.5}) is bounded above by

\begin{eqnarray}
\lefteqn{
\mathbb{E}\left\vert\mathbb{E}\left(\exp\left(iu\left[\zeta_n\sum_{k=0}^m a_k (w_{n+k}-\theta_n)\right]+iu w_{n-1}X_{n-1}\right)\Big/\mathcal{D}_j,j\neq n\right)\right\vert}\nonumber\\
&&=\mathbb{E}\left\vert\mathbb{E}\left(\mathbb{E}\left(\exp\left(iu\left[\zeta_n\sum_{k=0}^m a_k (w_{n+k}-\theta_n)\right]+i u w_{n-1}X_{n-1}\right)\Big/\mathcal{D}_{j,n},j\neq n\right)\Big/\mathcal{D}_j,j\neq n\right)\right\vert \nonumber\\
&&=\mathbb{E}\left\vert\mathbb{E}\left(\exp\left(iu \left[\zeta_n\sum_{k=0}^ma_k(w_{n+k}-\theta_n)\right]\right)\mathbb{E}\left(\exp\left(iu w_{n-1}X_{n-1}\right)\Big/\mathcal{D}_{j,n},j\neq n\right)\Big/\mathcal{D}_j,j\neq n\right)\right\vert \nonumber\\
&&=
\mathbb{E}\left\vert\Psi_\zeta\left(u\sum_{k=0}^ma_k(w_{n+k}-\theta_n)\right)\Psi_w\left(uX_{n-1}\right)\right\vert
\nonumber\\
&&\longrightarrow \mathbb{E}\left\vert\Psi_w\left(uX_1\right)\right\vert\mathbb{E}\left\vert\Psi_\zeta(uZ_n)\right\vert\quad\textrm{as }m\to\infty, \label{7.2+1}
\end{eqnarray}
where
$$
Z_n=\sum_{k=0}^\infty a_k(w_{k+1}-\theta_n).
$$
The preceding statement is valid in view of the dominated convergence theorem   combined with
the  use of  the well known fact that if $X$ and $Y$ are independent,  $h(.,.)$ is a bounded  function and $g(y)=\mathbb{E}(h(X,y))$ then $\mathbb{E}(h(X,Y)\vert Y)=g(Y)$.
\\
Now, if the first part of assumption $(i)$ of Theorem \ref{Edgeworth Expansion} holds then, since for all $n\geq 1$  $Z\neq 0~ a.s.$,  for $\vert u\vert>b$ for a given $b$, we let $\delta>0$ and $\epsilon>0$  be such that $P(\vert u Z_n \vert>\epsilon)>1/2$ and $\vert \Psi_{\zeta} (uZ_n)\vert1_{\vert u Z_n\vert>\epsilon}\le1-2\delta$. Then we have
\begin{eqnarray*}
\mathbb{E}\left\vert\Psi_{\zeta}(uZ_n)\right\vert&=&\mathbb{E}\big( \left\vert\Psi_{\zeta}(uZ_n)\right\vert   1_{\vert uZ_n\vert>\epsilon}\big)+\mathbb{E}\big(\left\vert\Psi_{\zeta}(uZ_n)\right\vert1_{\vert u Z_n\vert\le\epsilon}\big)\\
&\le&(1-2\delta)P(\vert u Z_n\vert>\epsilon)+1-P(\vert u Z_n\vert>\epsilon)=1-2\delta P(\vert uZ_n\vert>\epsilon)<1-\delta.
\end{eqnarray*}
In summary, for any $b>0$ there exists $\delta>0$ such that for all $\vert u\vert> b$, we have $\mathbb{E}(\vert\Psi_{\zeta}(uZ_n)\vert)<1-\delta$. Therefore for $m$ sufficiently large,  for all $\vert u\vert>b$, we have
$$
\mathbb{E}\left\vert\Psi_{\zeta}\left(u\sum_{k=0}^m a_k w_{k+1}\right)\right\vert\le1-\delta=e^{-\ell},
$$
for some $\ell>0$, which completes the proof of (\ref{new2.5}). We note that in our proof for (\ref{new2.5}),  without loss of generality, we can take $b=\ell$ (possibly by having to work with $\zeta_j/\ell$ instead of $\zeta_j$) to be in the same context as in  (2.5) of  G\"{o}tze and Hipp \cite{Gotze and Hipp}. This means that (2.5) of G\"{o}tze and Hipp  \cite{Gotze and Hipp} holds true under the first part of of assumption $(i)$ of Theorem \ref{Edgeworth Expansion}.
\\
Alternatively, when the second part of the assumption $(i)$ of Theorem \ref{Edgeworth Expansion} holds, we show (2.5) of G\"{o}tze and Hipp  \cite{Gotze and Hipp} continues to hold true. Recalling  that in this case the weights satisfy the Cramer condition rather than the innovations, to establish (2.5) of G\"{o}tze and Hipp  \cite{Gotze and Hipp} in this case,  we first note that the left hand side of (\ref{7.2+1}), i.e., $\mathbb{E}\left\vert\Psi_\zeta\left(u\sum_{k=0}^m a_k(w_{n+k}-\theta_n)\right)\Psi_w\left(uX_{n-1}\right)\right\vert
$, is bounded above, uniformly in $m\geq 1$, by $E|\Psi_{w}(u X_1)|$. Since in this case we have $X_1\neq 0~ a.s.$, for given $b$ and for  $\vert u\vert>b$, we let $\delta>0$ and $\epsilon>0$  be such that $P(\vert u X_1\vert>\epsilon)>1/2$ and $\vert \Psi_{w} (u X_1)\vert1_{\vert u X_1\vert>\epsilon}\le1-2\delta$. Then, we have
\begin{eqnarray*}
\mathbb{E}\left\vert\Psi_{w}(u X_1)\right\vert&=&\mathbb{E}\big( \left\vert\Psi_{w}(u X_1)\right\vert   1_{\vert u X_1\vert>\epsilon}\big)+\mathbb{E}\big(\left\vert\Psi_{w}(u X_1)\right\vert1_{\vert u X_1\vert\le\epsilon}\big)\\
&\le&(1-2\delta)P(\vert u X_1\vert>\epsilon)+1-P(\vert u X_1\vert>\epsilon)=1-2\delta P(\vert u X_1\vert>\epsilon)<1-\delta.
\end{eqnarray*}
The preceding relation implies that, for any $b>0$ there exists $\delta>0$ such that for all $\vert u\vert> b$, we have $\mathbb{E}(\vert\Psi_{w}(u X_1)\vert)<1-\delta$, which implies that  (2.5) of G\"{o}tze and Hipp  \cite{Gotze and Hipp} holds true under the second  part of of assumption $(i)$ of Theorem \ref{Edgeworth Expansion}.
\\
Finally, in our context, condition (2.6) of G\"{o}tze and Hipp \cite{Gotze and Hipp}
is clearly satisfied with our  choice of $\mathcal{D}_j=\sigma(w_{j-1},\zeta_j)$ in view of the Dynkin's $\pi-\lambda$ theorem (cf., e.g.,  Billingsley \cite{Bill}).
\\
Now the proof of Theorem \ref{Edgeworth Expansion} is complete. $\square$
\\
We note that the preceding proof for Theorem \ref{Edgeworth Expansion} implies the validity of   (\ref{EE}) when $\E|\zeta_1|^5<\infty$ and $\E|w_1|^5<\infty$, and also that of (\ref{newEE}) on assuming $\E|\zeta_1|^4<\infty$ and $\E|w_1|^4<\infty$, provided that in both cases assumptions $(i)$ and $(ii)$ of Theorem \ref{Edgeworth Expansion} hold.

\subsection*{Proof of Theorem \ref{CLT}}
Our  arguments below to prove  this theorem are  valid for both i.i.d. and symmetric multinomial weights as specified in Theorem \ref{CLT}. We also note that we only give the proof   for the conditional statement (\ref{CLT Conditional}) as it implies the unconditional statement  (\ref{CLT Unconditional}), in view of the dominated convergence theorem.
\\
For the use in the  proof of this theorem, as well as   in the proof of Theorem \ref{CLT G^*^stu} we define
\begin{eqnarray}
K          &:=&\underset{n\to\infty}{\lim} \E(w_1-\theta_n)^2, \label{K^prime-1} \\
K^{\prime} &:=&\underset{n\to\infty}{\lim}\E\Big((w_1-\theta_n)(w_2-\theta_n)\Big). \label{K^prime}
\end{eqnarray}
We note that, by virtue of Corollary \ref{Corollary 1}
$K$ and $K^{\prime}$ are  positive constants and $K^{\prime}<K$.
\\
For the ease of the notation, without loss of generality we assume that $\mu=0$ and also define
\begin{eqnarray*}
\mathfrak{R}_{n,w}&:=&\gamma_0\sum_{i=1}^n(w_i-\theta_n)^2+2\sum_{h=1}^n\gamma_h\sum_{j=1}^{n-h}(w_j-\theta_n)(w_{j+h}-\theta_n)\\
&=&\V \Big(\sum_{i=1}^n\big(w_i-\theta_n)X_i  \big|  w_{1}, \cdots, w_{n}  \Big),
\end{eqnarray*}
In view of   Theorem 2.2  of Abadir {\it et al.}  (2014), the proof of Theorem \ref{CLT} will result if we show that as $n\to\infty$,
\begin{equation}\label{ratio1}
\mathfrak{R}_{n,w}\big/ (n \mathfrak{D}_{n,X,w})- 1=o_{P_w}(1),
\end{equation}
\begin{equation}\label{ratio2}
\underset{1\le i\le n}{\max} \vert w_i-\theta_n \vert \big/\sqrt{n \mathfrak{D}_{n,X,w}}=o_{P_w}(1),
\end{equation}
and
\begin{equation}\label{ratio3}
\sum_{i=1}^n (w_i-\theta_n)^2 \big/ (n \mathfrak{D}_{n,X,w})=O_{P_w}(1).
\end{equation}
We remark that our condition (\ref{ratio3}) relates to condition $(ii)$ of Theorem 2.2 of   Abadir \textsl{et al}.  (2014). The latter condition which is intended for constant weights for linear processes reads as follows: $\exists$ $C>0$ such that

\begin{equation}\nonumber\label{abadir original condition}
\frac{1}{\sigma^2}\ \underset{n\geq 1}{\sup}\sum_{j=1}^n z^2_{nj}\le C,
\end{equation}
where $z_{nj}$ are  non-random  weights, $\sigma^2=\lim_{n \to \infty} \sum_{i=1}^n z_{nj} X_i$ and  $X_i$, $1\leq i \leq n$,  are the first $n$ terms of the linear process  (\ref{eq 1}).   The preceding condition can, conveniently and equivalently,   be replaced  by

\begin{equation}\label{abadir}
\frac{1}{\sigma^2}\sum_{j=1}^n z^2_{nj}=O(1),\qquad\textrm{as }n\to\infty.
\end{equation}
Hence, in the stochastic context of our  Theorem \ref{CLT},  the deterministic statement (\ref{abadir original condition}) is replaced by condition (\ref{ratio3}).
\\
\\
We now  establish (\ref{ratio1}) by writing

\begin{eqnarray}\label{ratio12}
&&\mathfrak{R}_{n,w}\big/ (n \mathfrak{D}_{n,X,w})-1\nonumber\\
&&=\frac{\gamma_0\frac{n^{-2d}}{n}\sum_{i=1}^n(w_i-\theta_n)^2+
2\frac{n^{-2d}}{n}\sum_{h=1}^n\gamma_h\sum_{j=1}^{n-h}(w_j-\theta_n)(w_{j+h}-\theta_n)}
{\gamma_0n^{-2d}\E(w_1-\theta_n)^2+2n^{-2d}\E[(w_1-\theta_n)(w_2-\theta_n)]
\sum_{h=1}^n\left(1-\frac{h}{n}\right)\gamma_h}-1\nonumber\\
&&=\Bigg[\gamma_0 n^{-2d}\left(\frac{\sum_{i=1}^n(w_i-\theta_n)^2}{n}-\E(w_1-\theta_n)^2\right)\nonumber\\
&&+2n^{-2d}\sum_{h=1}^n\gamma_h\left(\frac{1}{n}\sum_{j=1}^{n-h}(w_j-\theta_{n})(w_{j+h}-\theta_{n})-
\E\big((w_1-\theta_{n})(w_2-\theta_{n})\big)\right)\Bigg]\times\nonumber\\
&&\quad \left[\gamma_0 n^{-2d}\E(w_1-\theta_{n})^2+2 n^{-2d}\E\big( (w_1-\theta_n)(w_2-\theta_{n})\big)
\sum_{h=1}^n(1-\frac{h}{n})\gamma_h\right]^{-1}.\nonumber\\
\end{eqnarray}
Considering that, as $n\to\infty$, for i.i.d. and symmetric multinomial weights we have

\begin{eqnarray}\label{ratio13}
\lefteqn{
 \gamma_0 n^{-2d}\E(w_1-\theta_n)^2+2n^{-2d}\E\Big((w_1-\theta_{n})(w_2-\theta_{n})\Big)
\sum_{h=1}^n (1-\frac{h}{n})\gamma_h}\nonumber\\
&&\to\begin{cases}\gamma_0 (K -K^{\prime})+K^{\prime} s_X^2>0,&\textrm{ when }d=0,\\\\
K^{\prime}s_X^2>0,&\textrm{ when }0<d<1/2,\end{cases}
\end{eqnarray}
where
$$
s^{2}_X:=\lim_{n\to +\infty} \V \big( n^{1/2 -d} \bar{X}_n \big)=\lim_{n\to +\infty} n^{-2d} \big\{ \gamma_0+2\sum_{h=1}^{n-1} \gamma_h (1-h/n)\big\}.
$$
From  (\ref{ratio13}), the relation (\ref{ratio1}) will follow if we show that, as $n\to\infty$,

\begin{equation}\label{ratio14}
\frac{1}{n}\sum_{i=1}^n(w_i-\theta_n)^2-\E(w_1-\theta_n)^2=o_{P_w}(1)
\end{equation}
and
\begin{equation}\label{ratio15}
n^{-2d}\sum_{h=1}^n\gamma_h\frac{1}{n}\sum_{j=1}^{n-h}\Big((w_j-\theta_n)(w_{j+h}-\theta_n)-
\E \big( (w_1-\theta_n)(w_2-\theta_n)\big)\Big)\\
=o_{P_w}(1).
\end{equation}
When the weights are i.i.d.,  (\ref{ratio14}) is a consequence of the law of large numbers for row-wise i.i.d. triangular arrays of random variables, in view of the assumption that   $  \E|w_{1}|^{3}<\infty$, as in scheme {\textsf{(\textbf{RS})}} and also in Theorem \ref{CLT}.
\\
We now  prove that (\ref{ratio14}) continues to  hold  for the case when
\\
$(w_1,\ldots,w_n)\overset{d}{=} \mathcal{M}ultinomial(n;1/n,\cdots,1/n)$, by writing
\begin{eqnarray}
&&P\Big(\big\vert\frac{1}{n}\sum_{i=1}^n \big((w_i-\theta_n)^2-\E(w_1-\theta_n)^2\big)\big\vert>\epsilon\Big)\nonumber\\
&&\le\epsilon^{-2} \frac{n}{n^2} \E\left((w_1-\theta_n)^2-\E(w_1-\theta_n)^2\right)^2\nonumber\\
&&+\epsilon^{-2}\frac{n(n-1)}{n^2}\E\left[\left((w_1-\theta_n)^2-\E(w_1-\theta_n)^2\right)
\left((w_2-\theta_n)^2-\E(w_2-\theta_n)^2\right)\right].\nonumber\\
\label{add proof1}
\end{eqnarray}
The  first expectation above on the right hand side  of (\ref{add proof1}) is $O(1)$, as $n \to \infty$. This conclusion  is a consequence of conditions $(i)$  of Theorem \ref{CLT}. To show the asymptotic negligibility of the right hand side  of (\ref{add proof1}), it remains to show that the  second expectation in it   is $o(1)$, as $n \to \infty$. To do so we write,
\begin{eqnarray}
&&\E\left[\left((w_1-\theta_n)^2-\E(w_1-\theta_n)^2\right)
\left((w_2-\theta_n)^2-\E(w_2-\theta_n)^2\right)\right] \nonumber\\
&&=\E\Big((w_1-\theta_n)^2(w_2-\theta_n)^2\Big)-\E^2(w_1-\theta_n)^2 \nonumber\\
&&=\E\Big( (w_1w_2)^2-2 (w_1)^2 w_2+\theta_n^2 (w_1)^2-2\theta_n w_1(w_2)^2
+4\theta_n^2 w_1w_2-2 \theta_n^3 w_1 \nonumber\\
&&\ \ +\theta_n^2 (w_2)^2
-2\theta_n^3 w_2+\theta_n^4\Big)-\E^2(w_1)^2-4\E(w_1)\theta_n^2-\theta_n^4
+4\E^2(w_1)^2 \E(w_1)\theta_n \nonumber\\
&&\ \ -2  \E^2\left(w_1\right)+4\theta_n^3\E\left(w_1\right)\nonumber\\
&&\sim\left(4-4\theta^{*}+2(\theta^{*})^2-4 \theta^{*}+4(\theta^{*})^2-2(\theta^{*})^3+2 (\theta^{*})^2-2 (\theta^{*})^3+(\theta^{*})^4\right)\nonumber\\
&&\ \ -4-4(\theta^{*})^2-(\theta^{*})^4+8 \theta^{*} - 4 (\theta^{*})^2+4(\theta^{*})^3\nonumber\\
&&=0.\label{add proof2}
\end{eqnarray}
Therefore, (\ref{ratio14}) holds true for symmetric  multinomial weights too.
\\
As for (\ref{ratio15}), when weights are i.i.d.,   for all $\epsilon>0$, we have

\begin{eqnarray}\label{ratio16}
&&P\left(n^{-2d}\Big\vert\sum_{h=1}^n\gamma_h\frac{1}{n}\sum_{j=1}^{n-h}\Big((w_j-\theta_n)(w_{j+h}
-\theta_n)- \E\big( (w_1-\theta_n)(w_2-\theta_n)\big)\Big)\Big\vert>\epsilon\right)\nonumber\\
&&\le\epsilon^{-1}n^{-2d}\sum_{h=1}^n\vert\gamma_h\vert
\E^{1/2}\left(\frac{1}{n}\sum_{j=1}^{n-h}\Big((w_j-\theta_n)(w_{j+h}
-\theta_n)- \E\big( (w_1-\theta_n)(w_2-\theta_n)\big)\Big)\right)^2.\nonumber\\
\end{eqnarray}
Now observe  that for all $h\ge1$,
\begin{eqnarray*}
&&\E\left(\frac{1}{n}\sum_{j=1}^{n-h}\Big((w_j-\theta_n)(w_{j+h}
-\theta_n)- \E\big( (w_1-\theta_n)(w_2-\theta_n)\big)\Big)\right)^2\\
&&=\frac{1}{n^2}\sum_{j=1}^{n-h}
\E\left( (w_j-\theta_n)(w_{j+h}-\theta_n)-\E^2(w_1-\theta_n)\right)^2\\
&&=\frac{n-h}{n^2}\Big(\E\big( (w_1-\theta_n)(w_2-\theta_n)\big)^2-\E^4(w_1-\theta_n)\Big)
\\
\end{eqnarray*}
Substituting  the preceding relation into the right hand side of (\ref{ratio16}), in view of conditions $(i)$ and $(ii)$ of this theorem,  we conclude that
\begin{eqnarray*}
&&\epsilon^{-1}n^{-2d}\sum_{h=1}^n\vert\gamma_h\vert\frac{\sqrt{n-h}}{n}\Bigg(\E^2\left((w_1-\theta_n)^2\right)-\E^4(w_1-\theta_n)\Bigg)^{1/2}
\\
&&=o(1), \ \textrm{as}\ n\to\infty.
\end{eqnarray*}
The preceding convergence to zero holds uniformly in $h$ since
$\sup_{1\leq h \leq n-1} \sqrt{n-h}/n=\sqrt{n-1}/n\to 0$ and
$\displaystyle{n^{-2d}\sum_{h=1}^\infty\vert\gamma(h)\vert}$
is bounded.
Now the proof
of (\ref{ratio15}) in the case of i.i.d. weights is complete.
\\
The proof of (\ref{ratio15}) when  the weights are symmetric multinomial  also begins with the inequality (\ref{ratio16}). In this case, in view of condition $(i)$, as $n \to \infty$, we have

\begin{equation}\label{ratio16+1}
\E\Big( (w_{1}-\theta_n) (w_{2}-\theta_n)  \Big)^2 - \E^2\Big( (w_{1}-\theta_n) (w_{2}-\theta_n) \Big)=O(1),
\end{equation}

\begin{eqnarray}
&&\E\Bigg\{  \Bigg( (w_{1}-\theta_n) (w_{2}-\theta_n)- \E\Big( (w_{1}-\theta_n) (w_{2}-\theta_n) \Big) \Bigg)\nonumber\\
&& \qquad \times  \Bigg( (w_{3}-\theta_n) (w_{4}-\theta_n)- \E\Big( (w_{3}-\theta_n) (w_{4}-\theta_n) \Big) \Bigg)   \Bigg\}\nonumber\\
&&\sim \Big\{  1-2 \theta^{*}+(\theta^{*})^2- 2 \theta^{*} +4(\theta^{*})^2 -2(\theta^{*})^3   +(\theta^{*})^2
-2 (\theta^{*})^3 +(\theta^{*})^4 \nonumber \\
&& ~ ~ -1 -4 (\theta^{*})^2 -(\theta^{*})^4   +4 \theta^{*} -2(\theta^{*})^2 +4(\theta^{*})^3\Big\}\nonumber\\
&&= 0.\label{ratio16+2}
\end{eqnarray}
These last two approximations imply that the right hand side  of (\ref{ratio16}) is, asymptotically in $n$ and uniformly in $h$, negligible which means that (\ref{ratio15}) holds for symmetric multinomial weights. This completes the proof of (\ref{ratio1}).
\\
We now give a unified argument for both i.i.d. and symmetric multinomial weights to  prove (\ref{ratio2}). The proof is also valid for  both short and long memory data and it begins with  observing that, as $n \to \infty$, we have
\begin{equation}\label{ratio21}
\frac{1}{\sqrt{n}}\underset{1\le j\le n}{\max}\left\vert w_j-\theta_n\right\vert=o_{P_w}(1).
\end{equation}
Considering that $\theta_n$ is bounded, the weight $w_{j}$, for each $n$,  are identically distributed and, since,  in the scheme {\textsf{(\textbf{RS})}} the weights are so that $\sup_{n\geq 1} \E|w_{1}|^3<\infty$,   the proof of (\ref{ratio21})  follows from the following argument.

\begin{eqnarray*}
P\big(\underset{1\le j\le n}{\max}(w_j )^2>n \epsilon \big)&\le&
n P \big((w_1)^2>n\epsilon \big)\\
&\le& \epsilon^{-3/2} n^{-1/2} \E|w_{1}|^3
\le \epsilon^{-3/2} n^{-1/2}  \E|w_{1}|^3 \to 0.
\end{eqnarray*}
In view of the conditions $(i)$ and $(ii)$ of Theorem \ref{CLT}, for some finite number $A$,   we have
\begin{eqnarray*}\lefteqn{
\E(w_1-\theta_n)^2\gamma_0+2\E \Big( (w_1-\theta_n)(w_2-\theta_n)\Big)
\sum_{h=1}^n (1-\frac{h}{n})\gamma_h}\\
&&\longrightarrow K \gamma_0+2K^{\prime}\sum_{h=1}^\infty\gamma_h=\begin{cases}A,&\textrm{when }X_j\textrm{ of short memory},\\
\infty,&\textrm{when }X_j\textrm{ of long memory}.
\end{cases}
\end{eqnarray*}
Recall that  $K^{\prime} < K $. This, in turn,  implies that $A >0$. This together with  (\ref{ratio21}) completes the proof of (\ref{ratio2}).
\\
\\
To complete the proof of Theorem \ref{CLT}, we need to  prove (\ref{ratio3}). The validity of (\ref{ratio3}), for both i.i.d. and symmetric multinomial weights,  by virtue of
(\ref{ratio14})  and  conditions $(ii)$ of Theorem \ref{CLT}, results from the following weak law of large numbers, as $n \to \infty$.

\begin{eqnarray*}
&&\frac{(1/n)\sum_{j=1}^n(w_j-\theta_n)^2}{\E(w_1-\theta_n)^2\gamma_0+
2\E\Big((w_1-\theta_n)(w_2-\theta_n)\Big)
\sum_{h=1}^n (1-\frac{h}{n}) \gamma_h}\\
&&\overset{P_w}{\longrightarrow}\frac{K }{\gamma_0K +2K^{\prime}\sum_{h=1}^\infty\gamma_h}
=\begin{cases}A^{\prime} > 0,&\textrm{when }X_j\textrm{ of short memory},\\
0,&\textrm{when }X_j\textrm{ of long memory}.
\end{cases}
\end{eqnarray*}
This completes the proof of  (\ref{ratio3}) as well as that of  Theorem \ref{CLT}.  $\square$

\subsection*{Proof of Theorem \ref{CLT G^*^stu} }
In the proof of this    theorem, as well as in the proof of Theorem   \ref{Complete Studentization thm},     for the ease of the notation  we let $P_{X|w} (.)$ stand for the conditional probability $P(. \big| w_{1},\cdots,w_{n}  )$.
\\
As $n,q\to\infty$, in such a way that $q=O(n^{1/2})$, under the conditions of this theorem, from  the approximation  (2.10) of Theorem 2.1 of Abadir \textit{et al}. \cite{HACMAC}, that holds true for $d$ and $\hat{d}$,  we have

\begin{equation}\nonumber
q^{-2d}  \bar\gamma_0
+2q^{-2d} \sum_{h=1}
^q\bar\gamma_h(1-\frac{h}{q}) \longrightarrow s^{2}_X,
\end{equation}
where $s^{2}_X$ is as defined right after (\ref{ratio13}). This, in turn, implies that, as $n,q\to\infty$, in such a way that $q=O(n^{1/2})$, we also  have

\begin{eqnarray}\label{berno1}
q^{-2d}\mathcal{S}_{n,q,w}&=&q^{-2d}\E(w_1-\theta_n)^2 \bar\gamma_0
+2q^{-2d}\E\Big((w_1-\theta_n)(w_2-\theta_n)\Big)\sum_{h=1}
^q\bar\gamma_h(1-\frac{h}{q})\nonumber\\
&&\overset{P}{\longrightarrow}\begin{cases}\gamma_0(K -K^{\prime})+K^{\prime}s_X^2>0,&\textrm{ when }d=0,\\\\
K^{\prime} s_X^2>0,&\textrm{ when }0<d<1/2,
\end{cases}
\end{eqnarray}
where the constant $K$ is as in (\ref{K^prime-1}) and
$K^{\prime}$ is as in (\ref{K^prime}).
The preceding convergence means that  the Studentizing sequence $q^{-2d}\mathcal{S}_{n,q,w}$ asymptotically coincides with the limit of   $n \mathcal{D}_{n,X,w}$, which  is the   normalizing sequence for $T_{n,X,w}(\theta_n)$, as in (\ref{randomT}). In Theorem \ref{CLT} we showed $T_{n,X,w}(\theta_n)$ has standard  normal limiting distribution. Considering that

\begin{equation}\label{19+1}
G_{n,X,w}(\theta_n,d)= T_{n,X,w}(\theta_n) \sqrt{(n \mathcal{D}_{n,X,w})/(q^{-2d}\mathcal{S}_{n,q,w})},
\end{equation}
we conclude, from Slutsky theorem, that $G_{n,X,w}(\theta_n,d)$ also convergence to standard normal.
The relations (\ref{berno1}) and (\ref{19+1}) are also true when the memory parameter $d$ is replaced by its  estimator $\hat{d}$, provided that $\hat{d}-d=o_{p}(1/\log n)$.
\\
Now the  proof of Theorem  \ref{CLT G^*^stu} is complete. $\square$

\subsection*{Proof of Theorem  \ref{Complete Studentization thm}}
The proof of   Theorem \ref{Complete Studentization thm}, for both i.i.d. and symmetric multinomial weights, will follow if we show that

\begin{eqnarray}\label{berno12}
&& P_{X\vert w} \left\{ \big|\big( (q^{-2d} \hat{\mathcal{S}}_{n,q,w}) \big/ (q^{-2d}\mathcal{S}_{n,q,w})\big) -1 \big| > \epsilon \right\}\nonumber \\
 &=&P_{X\vert w}\left\{\left\vert\frac{q^{-2d}\bar\gamma_0\frac{1}{n}\sum_{j=1}^n(w_j-\theta_n)^2+2q^{-1-2d}\sum_{h=1}^q\bar\gamma_h\sum_{j=1}^{q-h}(w_j-\theta_n)(w_{j+h}-\theta_n)}
{q^{-2d}\bar\gamma_0\E(w_1-\theta_n)^2+2q^{-2d}\E\Big((w_1-\theta_n)(w_2-\theta_n)\Big)
\sum_{h=1}^q\bar\gamma_h(1-\frac{h}{q})}-1\right\vert>\epsilon\right\}\nonumber\\
&&=o_{P_w}(1),\qquad\textrm{as }n\to\infty,
\end{eqnarray}
where $\epsilon$ is an arbitrary positive constant. By virtue of (\ref{berno1}), the proof of (\ref{berno12}), as $n,q \to \infty$ such that $q=O(n^{1/2})$,  will follow from (\ref{ratio14}) combined with

\begin{equation}\label{berno12+1}
P_{X|w} \big( \big| q^{-2d} \sum_{h=1}^q\bar\gamma_h B_{n,q,w}(h) \big|> \epsilon  \big) =o_{P_w}(1),
\end{equation}
where
\begin{equation*}
B_{n,q,w}(h):= q^{-1} \sum_{j=1}^{q-h} \Big( (w_j-\theta_n)(w_{j+h}-\theta_n)
- \E \big((w_1-\theta_n)(w_2-\theta_n)\big) \Big).
\end{equation*}
The proof of (\ref{berno12+1})  is  a modification of the proof  of (6.14) in Cs\"{o}rg\H{o} \textit{et al.} \cite{Csorgo and Nasari and Oul-Haye}.
\\
For convenient reference here  we  give the proof of (\ref{berno12+1}) for both i.i.d. and symmetric multinomial weights. To do so, without loss of generality, we assume that $\mu=\E X_{1}=0$   and, for each $1\leq h\leq q$, we define

\begin{equation}\label{def. gamma*}
\gamma_{h}^*:=\frac{1}{n} \sum_{i=1}^{n-h}  X_{i} X_{i+h}.
\end{equation}
Observe   that, for $\varepsilon_1, \ \varepsilon_2>0$, we have
\begin{eqnarray}
&&P\big\{ P_{X|w}\big( q^{-2d}\big|  \sum_{h=1}^q \bar{\gamma}_{h} B_{n,q,w}(h)   \big|> 2\varepsilon_1   \big)>\varepsilon_2 \big\}\nonumber\\
&&\leq P\big\{ P_{X|w}\big( q^{-2d}\big|  \sum_{h=1}^q (\bar{\gamma}_{h}-\gamma_{h}^*) B_{n,q,w}(h)   \big|>\varepsilon_1   \big)>\varepsilon_2 \big\}\nonumber\\
&&\qquad +  P\big\{ P_{X|w}\big( q^{-2d} \big|  \sum_{h=1}^q \gamma_{h}^* B_{n,q,w}(h)   \big|>\varepsilon_1   \big)>\varepsilon_2 \big\}.\label{eq 21 proofs}
\end{eqnarray}
We now show that the first term in (\ref{eq 21 proofs}) is asymptotically negligible, noting first that
\begin{eqnarray}
\sum_{h=1}^q (\bar{\gamma}_{h}-\gamma_{h}^*) B_{n,q,w}(h)&=&-\frac{\bar{X}_n}{n} \sum_{h=1}^q  B_{n,q,w}(h)  \sum_{i=1}^{n-h} X_i - \frac{\bar{X}_n}{n} \sum_{h=1}^q  B_{n,q,w}(h) \sum_{i=1}^{n-h} X_{i+h}\nonumber\\
&&+
\bar{X}^2 \sum_{h=1}^q B_{n,q,w}(h)\nonumber\\
&\sim& - \bar{X}^2  \sum_{h=1}^q B_{n,q,w}(h)\ uniformly \ in \ h\ in \ probability-P_{X|w},\nonumber\\
&&\label{eq 22 proofs}
\end{eqnarray}
where, in the preceding conclusion, generically,  $Y_{n}\sim Z_{n}$ in probability-$P$ means $Y_n=Z_n (1+o_{P}(1))$. The approximation in (\ref{eq 22 proofs}) is true since,  for $\varepsilon>0$ we have
\begin{eqnarray*}
P\big( \cup_{1\leq h \leq q} \big|  \bar{X}_{n} - \frac{\sum_{i=1}^{n-h} X_i}{n} \big|>\varepsilon \big)&\leq&q  P\big(  \big|  \frac{\sum_{i=n-h+1}^{n} X_i}{n} \big|>\varepsilon \big)\label{newly added 1}\\
&\leq& \varepsilon^{-4}  q \frac{(h-1)^4}{n^4} \E(X^{4}_1)\nonumber\\
&\leq&\varepsilon^{-4} \frac{q^5}{n^4} \E(X^{4}_1)\rightarrow 0,\ as \ n\to \infty.\nonumber
\end{eqnarray*}
The preceding is true since $1\leq h \leq q$ and $q=O(n^{1/2})$, as $n,q \to \infty$.
\\
We note that   for $0\leq d <1/2$, as $n\rightarrow \infty$, we have that  $n^{1/2-d} \bar{X}_{n}=O_{P}(1)$. The latter conclusion, in view of  the equivalence in (\ref{eq 22 proofs}), implies that, for each $\varepsilon_1, \varepsilon_2>0$, there exists   $\varepsilon>0$ such that
\begin{eqnarray}
 &&P\big\{ P_{X|w}\big( q^{-2d} \big|  \sum_{h=1}^q (\bar{\gamma}_{h}-\gamma_{h}^*) B_{n,q,w}(h)   \big|>\varepsilon_1   \big)>\varepsilon_2 \big\}\nonumber\\
&\sim& P \big\{ \frac{q^{-2d}}{n^{1-2d}} \sum_{h=1}^q \big|B_{n,q,w}(h)\big|>\varepsilon\big\}\nonumber \\
&\leq& \varepsilon^{-1} \frac{q^{-2d}}{n^{1-2d}} \sum_{h=1}^q \E\big(\big|B_{n,q,w}(h)\big|\big). \label{change 1}
\end{eqnarray}
From condition $(ii)$ of Theorem \ref{CLT}, that is also assumed in Theorem \ref{CLT G^*^stu}, there exists a constant $\mathcal{L}$ whose value does not depend on $n$ such that   $\sup_{n\geq 2}\sup_{1\leq h\leq q} \E \big(\big|B_{n,q,w}(h)\big|\big) < \mathcal{L}$. Hence,  (\ref{change 1}) can be bounded  above by
\begin{equation*}
\mathcal{L} \ \varepsilon^{-1}  \frac{q^{1-2d}}{n^{1-2d}}\longrightarrow 0,
\end{equation*}
as $n,q \to \infty$ in such away that $q=O(n^{1/2})$. This means that the first term in (\ref{eq 21 proofs}) is asymptotically negligible.  To establish   (\ref{berno12}) we show that   the second term in (\ref{eq 21 proofs}) is also asymptotically negligible.  To prove this negligibility, we  first define
\begin{equation}\label{def. of gamma**}
\gamma_{h}^{**}:=\frac{1}{n} \sum_{i=1}^n X_i X_{i+h}.
\end{equation}
Now, observe that
\begin{eqnarray}
P\big\{  \cup_{1\leq h \leq q} |\gamma_{h}^{**}-\gamma_{h}^{*}|>\varepsilon \big\}&\leq& q P\big\{   \frac{1}{n}| \sum_{i=n-h+1}^n X_i X_{i+h}|>\varepsilon \big\}\nonumber\\
&\leq& \varepsilon^{-2} \frac{q^3}{n^2} \E (X^{4}_1)\to 0,\nonumber
\end{eqnarray}
as $n,q\to \infty$ such that $q=O(n^{1/2})$,  hence, as $n,q\to \infty$ such that $q=O(n^{1/2})$, using an argument  similar  to those used for    (\ref{eq 21 proofs}) and (\ref{change 1}),  with $\gamma^{*}_h$ replacing $\bar{\gamma}_h$ and $\gamma^{**}_h$ replacing $\gamma^{*}_h$ therein,  we  arrive at
\begin{eqnarray*}
&&P\big\{ P_{X|w}\big( q^{-2d}\big|  \sum_{h=1}^q \gamma_{h}^{*} B_{n,q,w}(h)   \big|>\varepsilon_1   \big)>\varepsilon_2 \big\}\\
&\sim& P\big\{ P_{X|w}\big( q^{-2d}\big|  \sum_{h=1}^q \gamma_{h}^{**} B_{n,q,w}(h)   \big|>\varepsilon_1   \big)>\varepsilon_2 \big\}.
\end{eqnarray*}
Therefore, in order to prove (\ref{berno12}), it suffices to show that, as $n,q\to \infty$ so that $q=O(n^{1/2})$,
\begin{equation*}\label{eq 36 proofs}
P\big\{ P_{X|w}\big( q^{-2d}\big|  \sum_{h=1}^q \gamma_{h}^{**} B_{n,q,w}(h)   \big|>\varepsilon_1   \big)>\varepsilon_2 \big\}\to 0,
\end{equation*}
The  preceding  relation, in turn,   follows from the following two conclusions: as $n,q \to \infty$ so that $q=O(n^{1/2})$,

\begin{equation}\label{eq 37 proofs}
\sup_{1\leq h,h^\prime \leq q}  \E\big( \big|B_{n,q,w}(h) B_{n,q}(h^{\prime})  \big| \big)=o(1)
\end{equation}
and
\begin{equation}\label{eq 38 proofs}
q^{-4d} \sum_{h=1}^q \sum_{h^{\prime}=1}^q  \big| \E(\gamma_{h}^{**}  \gamma_{h^{\prime}}^{**}) \big|=O(1).
\end{equation}
To prove (\ref{eq 37 proofs}),  using the Cauchy inequality we  write

\begin{eqnarray}
&&\E\big( \big|B_{n,q,w}(h) B_{n,q,w}(h^{\prime})  \big| \big)\nonumber\\
&\leq& \E\big( B_{n,q,w}(h) \big)^2 \nonumber\\
&\leq&   \frac{q-h}{q^2} \ \E\Big( (w_{1}-\theta_n) (w_{2}- \theta_n)    - \E \big( (w_{1}- \theta_n) (w_{2} - \theta_n) \big)  \Big)^2 \nonumber \\
&+& \frac{(q-h)(q-h-1)}{q^2}   \E \Bigg\{ \Big(   (w_{1}- \theta_n) (w_{2}-\theta_n) - \E \big( (w_{1}- \theta_n) (w_{2}-\theta_n) \big) \Big) \nonumber\\
 && ~~~~~~~~~~~~~~~~~~~~~~~~~ \times \Big( (w_{3}- \theta_n) (w_{4}- \theta_n)- \E \big( (w_{3}- \theta_n) (w_{4}- \theta_n) \big)    \Big)  \Bigg\}. \nonumber \\
 \label{added eqq1}
\end{eqnarray}
In case of      i.i.d. weights, (\ref{added eqq1}) is bounded above by $q^{-1} \E^{2} (w_1 - \theta_n)^2$ that
vanishes as $q, n \to \infty$. In case of symmetric multinomial weights, from (\ref{ratio16+1}) and (\ref{ratio16+2}) we can see that (\ref{added eqq1}) has an  upper bound of the form $2 K_n/q + k_n$, where $K_n=O(1)$ and $k_n=o(1)$, that also vanishes as $n,q \to \infty$. The latter conclusion completes the proof of (\ref{added eqq1}) and  that of (\ref{eq 37 proofs}).
\\
In order to establish (\ref{eq 38 proofs}),  we define
 \begin{equation*}
H:=\lim_{s\to \infty} s^{-2d} \sum_{\ell =-s}^{s} |\gamma_{\ell}|.
\end{equation*}
Observe that $H<\infty$.
We now carry on with the proof of (\ref{eq 38 proofs}),  using  a generalization of an argument used in the proof of Proposition 7.3.1 of Brockwell  and Davis \cite{Brockwell and Davis} as follows:

\begin{eqnarray}
q^{-4d}\sum_{h=1}^q  \sum_{h^\prime=1}^q \big|  \E(\gamma_{h}^{**}  \gamma_{h^{\prime}}^{**}) \big|
 &\leq& q^{-2d}\sum_{h=1}^q  \big|\gamma_h \big|\ q^{-2d} \sum_{h^\prime=1}^q \big|\gamma_{h^\prime}|\nonumber\\
&+&(\frac{q}{n})^{1-2d} n^{-2d} \sum_{k=-n}^n \big| \gamma_h \big|\ q^{-2d} \sum_{L=-q}^q  \big| \gamma_{k+L} \big|\nonumber \\
&+& \frac{1}{n} \sum_{k=-n}^n \ q^{-2d} \ \sum_{h^\prime=1}^q  \big| \gamma_{k+h^\prime} \big| \ q^{-2d} \ \sum_{h=1}^q \big| \gamma_{k-h} \big|\nonumber\\
&+& \frac{q^{-2d}}{n^{1-2d}}  n^{-2d} \sum_{i=1}^n \sum_{k=-n}^n |a_i a_{i+k}|\  q^{-d} \sum_{h=1}^q |a_{i+h}| \ q^{-d} \sum_{h^\prime=1}^q |a_{i+k-h^\prime}|.\label{eq 39 prroofs}
\end{eqnarray}
It is easy to see that, as $n\to \infty$, and consequently $q\to \infty$, the right hand side of the inequality  (\ref{eq 39 prroofs}) converges to the finite limit $3 H^2$. Now the proof of (\ref{eq 38 proofs}) and also that of Theorem \ref{Complete Studentization thm} are complete. $\square$


\end{document}